\documentclass[aps,physrev,reprint,superscriptaddress]{revtex4-2}
\usepackage{ragged2e}
\usepackage{graphicx} 
\usepackage{amsmath}
\usepackage{physics}
\usepackage{float}
\usepackage{hyperref}
\usepackage{color}
\usepackage{amsfonts}
\usepackage{amssymb}
\usepackage{ragged2e}
\AtBeginDocument{
  \setlength{\abovedisplayskip}{3pt}
  \setlength{\belowdisplayskip}{3pt}
  \setlength{\abovedisplayshortskip}{2pt}
  \setlength{\belowdisplayshortskip}{2pt}
  \setlength{\jot}{2pt}
}
\makeatletter
\long\def\@makecaption#1#2{%
  \par
  \vskip\abovecaptionskip
  \begingroup
    \small\rmfamily
    \setlength{\parindent}{0pt}%
    \justifying
    \let\\\@normalcr
    \let\footnote\@footnotemark@gobble
    \@make@capt@title{#1}{#2}\par
  \endgroup
  \vskip\belowcaptionskip
}
\makeatother

\begin{document}


\title{Everything is a ``Spin'': The Secret Lives of SU(2)}


\author{Avik W. Ghosh}
\affiliation {Department of Physics, University of Virginia, Charlottesville, VA 22904, USA}
\affiliation{Department of Electrical and Computer Engineering, University of Virginia, Charlottesville,VA-22904, USA}


\date{\today}

\begin{abstract}
Spin, pseudospin, valley, polarization, and other two-component degrees of freedom share the geometry of SU(2), yet their topological manifestations are usually discussed as separate phenomena. This review develops a common geometric language for their winding in momentum and real space, beginning with Berry phase and the Dirac Hamiltonian and extending to graphene, topological insulators, Weyl semimetals, and magnetic skyrmions. We argue that the common thread is not merely topology itself, but the continuity constraints imposed on two-component wavefunctions. Whenever the relevant symmetry is preserved, winding determines which states can continuously connect across an interface or deformation, thereby governing transmission, torque generation, optical selection rules, and other physical responses. We then ask a practical question: what does topology buy an engineer? In skyrmions, winding partitions magnetic configuration space and stabilizes ultrasmall information carriers with tunable dynamics. In graphene, pseudospin matching governs Klein tunneling, enabling a gate-controlled transmission gap without sacrificing the massless Dirac dispersion. In topological insulators and Weyl semimetals, spin-momentum locking and Berry-curvature engineering generate electrically selectable spin currents, while helicity-dependent optical transitions produce circular photogalvanic responses. Together, these examples suggest that topology is not merely a classification of quantum matter, but a design language in which symmetry-protected wavefunction continuity can be engineered for memory, switching, actuation, and sensing.
\end{abstract}


\maketitle
\flushbottom

\section{The Topological connection between materials}
The goal of this paper is two-fold: 
(a) Examining the parallelism between spins, pseudospins, polarization and other ``SU(2)'' variants, specifically the topological similarity in the winding patterns of their wavefunctions; and (b) postulating device applications that uniquely arise from gating these symmetries.

Consider a $2 \times 1$ complex vector $(\alpha, \beta)^T$,
representing  mixing coefficients for up/down electron spins,
left/right circular polarization,
magnon or cavity modes,  or for that matter, 
any generic qubit. Mathematically these two-state systems are identical, with the same underlying geometry - that of rotations among the two amplitudes that preserve normalization $|\alpha|^2 + |\beta|^2$.
That geometry is called SU(2), for Special Unitary Group of $2\times 2$ unitary matrices $U$ of determinant unity that transform these spinors while preserving normalization. 
A normalized two-component complex state lies on the three-sphere 
$S^3$ embedded in a four-dimensional manifold.
Factoring out its physically irrelevant overall phase produces the Bloch sphere S$^2$ surface. 
For physical rotations, SU(2) double-covers the 3-D Special Orthogonal symmetry group SO(3): two spinors $\pm u$ correspond to the same Bloch vector, and a $2\pi$ rotation changes the spinor sign while a $4\pi$ rotation restores it.
The Pauli matrices $\vec{\sigma}$ generate the transformations $U = \exp{-i\theta\hat{n}\cdot\vec{\sigma}/2}$ and the corresponding Bloch vector rotates as an ordinary 3-D vector.

The central thesis of this write-up is the following: 
topological winding constrains how two SU(2) states continuously connect across an interface. Whenever symmetry preserves the relevant quantum number, these constraints directly govern transmission, deformation, torque generation, and optical selection rules. We argue that these geometric restrictions can be deliberately engineered into useful device functionality.
\begin{figure*}[ht] 
  \centering
\includegraphics[width=4.83in]{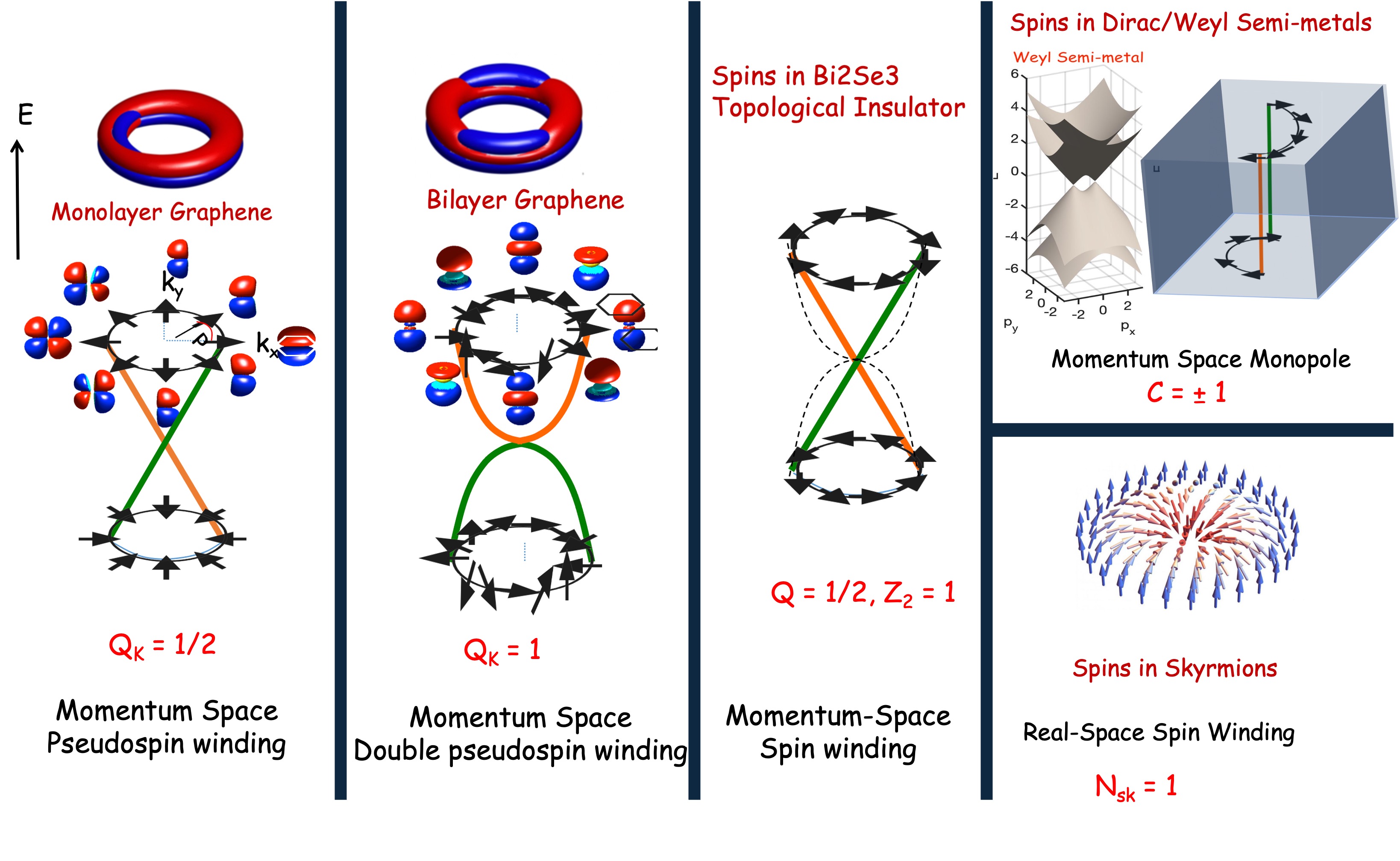} 
 \caption{Topological winding textures in momentum and real space are all manifestations of the same geometry spanning Bloch, Poincaré or valley pseudospin spheres. For monolayer graphene, the pseudospin winds once around a single valley, giving a valley-resolved winding charge $Q_K
=\gamma/2\pi=1/2$. The pseudospin describes the mixing coefficients of the laterally separated sublattice p$_z$ orbitals that constitute the corresponding eigenstate (red is positive, blue is negative).  
Bernal stacked bilayer graphene exhibits a double pseudospin winding, yielding $Q_K =1$. Here the pseudospin corresponds to the relative amplitudes of the low-energy non-dimer orbital states distributed across the two graphene layers. The top surface state of a 3D topological insulator likewise contains a single Dirac cone with Berry phase $\gamma = \pi$, corresponding to a half-integer winding charge $Q=1/2$; this isolated surface cone is protected by the nontrivial bulk $Z_2 =1$ topology. Weyl nodes act as Berry curvature monopoles with Chern charge $C=\pm 1$, while magnetic skyrmions are characterized by a real-space winding number $N_{sk}=1$.}
  \label{fig:wind}
\end{figure*}
\subsection{Apples and oranges}
What is the {\it{geometrical}} difference  between an apple and an orange? Topologically, they are identical, as we can deform one continuously into the other through a one-to-one mapping, unlike say, a donut. In fact, the distinction is loosely, the number of holes (more accurately, the maximum number of simple closed non-intersecting curves on its surface that do NOT divide the object into separate pieces). We can define a topological index $g$, the `genus', obtained by integrating its Gaussian curvature $K$ 
\begin{equation}
    \int KdS = 4\pi(1-g) ~~~({\rm{Gauss-Bonnet~theorem}})
\end{equation}
A sphere has Gaussian curvature $K = 1/R^2$ which leads to $g = 0$. A single closed line on its surface separates it. In contrast, a donut has positive curvature on the outer periphery and negative on the inside by the hole, integrating to zero, yielding $g = 1$. We can draw at most one closed curve without separating it into disjoint pieces. 

  Imagine a Flat-Lander (or a Flat-Earther) lying supine on a spherical surface  traversing a closed curve - say a triangle, by starting at a point, `parallel transporting' an arrow without rotating in the local tangent plane as it follows the contour of the sphere, then continuing along the second side similarly and back along a third side to the origin. What the Flat-Earther finds in the process is the arrow has rotated relative to its original orientation by an angle - the {\it{holonomy}} - that is the integral over the contour of the curvature. 
This thought experiment is actually executed by the Foucault pendulum in Paris, which swings along an inertial plane as the earth rotates below it taking the local normal and tangent plane along for a daily ride. After a full day, the plane of the pendulum will have rotated relative to the earth by an angle 
set by the integral of the curvature over the closed contour area, set by the latitude $\theta$
\begin{equation}
    \int KdS = 2\pi(1-\sin{\theta}) = -2\pi\sin{\theta}~~{\rm{mod}}(2\pi)
\end{equation}
Think of the curved earth above the latitude as a skull cap  made out of a circle drawn on a flat piece of paper. The circle has too much circumference per unit area and will need a wedge cut into it to fold into the cap. The missing wedge angle is the  holonomy $2\pi\sin{\theta}$. For a bowl the curvature is negative, and the holonomy is $-2\pi\sin{\theta}$ - the circle has too little circumference to cover it. 

At Paris, the Foucault precession modulo $2\pi$ corresponds to roughly 89$^0$, meaning that the pendulum will be seen to swing almost perpendicular to its original plane if we were to return in a day. Significantly, this diurnal swing depends on the integral of the curvature, not just the local value along the path traced by the pendulum. Curiously, it is a global property.

The swing in  phase is seen in condensed matter physics in the celebrated Aharonov-Bohm effect. An electron traveling along a closed path enclosing a solenoid picks up a Peierls phase set by the vector potential, $\vec{k} \rightarrow \vec{k} - q\vec{A}/\hbar$, the Peierls construction needed to get the correct Lorentz force. The total phase around the closed curve, using Stokes' theorem, is then $\oint \Delta \vec{k}\cdot d\vec{l} = -(q/\hbar)\oint \vec{A}\cdot d\vec{l} = -(q/\hbar)\int \vec{\nabla}\times\vec{A}\cdot d\vec{S} = -q\Phi_B/\hbar$, proportional to the enclosed magnetic flux. The electron `senses' the magnetic field from the global integrated curvature of the electromagnetic potential, even though the latter never penetrates outside the solenoid where the electron sits.

We can generalize this concept to any adiabatically curving Hamiltonian \cite{bernevig,vanderbilt2018}. Let $t$ parametrize a closed curve, say the time index in k-space, along which the eigenvalue $\epsilon_{\vec{k}(t)}$ varies continuously as $\vec{k}(t)$ evolves along the curve. The wavefunction will pick up a net phase $\gamma$
\begin{equation}
    \psi_{\vec{k}(t)} = e^{\displaystyle i\gamma(t)}e^{\displaystyle -i\int_0^t \epsilon_{\vec{k}(t^\prime)}dt^\prime/\hbar}\psi_{\vec{k}(0)}
\end{equation}
The Berry phase $\gamma$, obtained by substituting this wavefunction into the time-dependent Schr\"odinger equation, gives us the generalization of the Aharonov-Bohm phase
\begin{eqnarray}
    \gamma &=& \oint \vec{A}\cdot d\vec{k} = \int \vec{\Omega}_k\cdot d\vec{S}_k~~~({\rm{Berry~Phase}})\nonumber\\
    \vec{\Omega}_k &=& \vec{\nabla}_k \times \vec{A}_k ~~~({\rm{Berry~Curvature}})\nonumber\\
    {A}_i(\vec{k}) &=& i\langle u_k|\partial_{i}u_k\rangle ~~~({\rm{Berry~Connection}})
    \label{eq:berry}
\end{eqnarray}
where $\partial_i = \partial/\partial_{k_i}$, and $u_k$ is the Bloch part of the wavefunction. The Berry connection senses the winding of the wavefunction, 
resembling the vector potential of a solenoid, driven by the current density $J \propto i[u^*\nabla u - u\nabla u^*$], except here the current is in momentum space and the $u$ is normalized, which means the two terms in the bracket are negatives of each other.

The winding of a two-component state has both an orientational and a metric aspect. The quantum geometric tensor (QGT) \cite{qgt} is defined for a non-degenerate band as 
\begin{eqnarray}
Q_{ij}(k) &=&  
\langle \partial_i u|(1-|u\rangle \langle u|)|\partial_j u\rangle 
\nonumber\\
&=&g_{ij} -i\Omega_{ij}/2
\end{eqnarray}
The derivative $|\partial_iu\rangle$ contains a gauge-dependent phase change parallel to the occupied state and one orthogonal to it signifying a genuine change in the quantum state. The projector $1-|u\rangle\langle u|$ removes the former, leaving only the physically meaningful geometric deformation.
Its imaginary, antisymmetric part $\Omega_{ij} =-2Im(Q_{ij}) = -2Im\langle \partial_iu|\partial_ju\rangle$ 
gives the Berry curvature and records the phase accumulated by the spinor, whereas its real, symmetric part—the quantum metric $g$—measures the local quantum distance signifying how rapidly neighboring spinors become distinguishable, $|\langle u(\vec{k})|u(\vec{k}+d\vec{k}\rangle|^2 = 1 - g_{ij}dk_idk_j + \ldots$. Transport and optical responses therefore probe not only the integrated winding of the state, but also the local rate at which the underlying spin, pseudospin, or orbital texture changes in momentum space, both arising from the same neighboring-spinor overlaps.

The local geometry encoded in the QGT governs a broad class of measurable electronic response functions, extending well beyond topological materials. In conventional materials these responses depend on the momentum-space distribution of the Berry curvature and quantum metric, whereas in topological phases their Brillouin-zone integrals can acquire quantized invariants such as the Chern number. While the quantum metric governs quantities such as localization, superfluid weight, and optical oscillator strength, the Berry curvature gives rise to transverse transport phenomena. As one familiar example, the intrinsic transverse conductivity is an occupation-weighted integral of the Berry curvature,
\begin{equation}
    \sigma_{ij}^{int} = -\dfrac{q^2}{\hbar}\epsilon_{ijl}\sum_n\int_{\rm{BZ}}\dfrac{d^dk}{(2\pi)^d}f_{nk}\Omega_{nl}(\vec{k}) 
\end{equation}
A partially filled metal measures the distribution of Berry curvature over the occupied Fermi sea with Fermi function $f_n(k)$. For a 2-D insulator at zero temperature, all states in the occupied bands are filled, and the Brillouin-zone integral becomes the integer Chern number,
\begin{eqnarray}
    C_n &=& \dfrac{1}{2\pi}\int_{BZ}d^2k\Omega_{n,z} ~~(={{0,\pm 1,\pm 2, \ldots}})\nonumber\\
    \sigma_{xy} &=& \dfrac{q^2}{h}\sum_{n\in~ {\rm{occup}}}C_n
\end{eqnarray}
What makes it `topological' is its ability to support a non-zero Chern number, or an equivalent topological index. 

Let us discuss a few variants, using graphene's simple two-band model as a convenient playground. 

\subsection{What do electrons in graphene look like?}
The Brillouin zone for graphene is a hexagon with each triangle connecting next-nearest neighbors containing equivalent points separated by a lattice vector, giving us two inequivalent Brillouin zone points (K, K' valleys) arising from the two inequivalent sublattice p$_z$ basis set orbitals. 
The corresponding low-energy spinless Hamiltonian, in the sublattice p$_z$ basis, expanded around each Dirac point at the Brillouin zone edge, is \cite{castro2009}
\begin{eqnarray}
    H_0(k) &=& \hbar v_F(\tau_zk_x\sigma_x +k_y\sigma_y) \nonumber\\
    &=& \hbar v_F\left(\begin{array}{cc}0 & \pm k_x -ik_y \\ \pm k_x + ik_y & 0\end{array}\right)
\end{eqnarray}
where the Pauli matrices $\tau$ correspond to the K, K' valleys, while the Pauli matrix $\sigma$ refers to the electron sublattice type A vs B atoms. The Pauli matrix products are meant to be Kr\"onecker products $\otimes$ so $H_0$ is a $4\times 4$ matrix with 4 low-energy bands. The energy eigenvalues $E = \pm\hbar v_F\sqrt{k_x^2+k_y^2}$ around each Dirac point give zero rest mass and no constant parabolic band mass; its energy-dependent cyclotron mass is $m^* = p/v = \hbar k/v_F$, while the atomistic (Bloch) component of the eigenstates, i.e., mixing coefficients of the sublattice p$_z$ orbitals, generate the two-component spinors
\begin{equation}
    u_k = \dfrac{1}{\sqrt{2}}\left(\begin{array}{c}1 \\ \pm e^{\displaystyle i\theta_k}\end{array}\right), ~~~\theta_k = \tan^{-1}{(k_y/k_x)}
\end{equation}
with the signs flipping between CB and VB, and between K and K$^\prime$ valleys. For positive momenta in the CB, the coefficients describe the bonding combination of the two sublattice $p_z$ orbitals, while  for negative momenta, it is the anti-bonding state. These two states, being orthogonal spinors, are designated as `pseudospins', and wind continuously around the Brillouin zone (Fig.~\ref{fig:wind} panel 1). 

We can now compute the Berry phase for each valley from the gradient winding of the wavefunctions (Eq.~\ref{eq:berry}) 
\begin{equation}
    \gamma = \begin{cases}\mp \pi,~~ {\rm{K~valley~CB/VB}}\\ \pm\pi,~~ {\rm{K}}^\prime{\rm{~valley~CB/VB}}\end{cases}
\end{equation}
If we endow the bands with a low-energy Dirac mass by opening a gap and placing a Fermi energy mid-gap, the total Chern number obtained by integrating the Berry curvature over the occupied bands across the entire Brillouin zone (i.e. all valleys) will be an integer. In fact, the Chern number $C = [{\rm{sign}}(m_K)-{\rm{sign}}(m_{K^\prime})]/2$ up to valley/sign convention, so that flipping Dirac mass $m_{K,K^\prime}$ between valleys will make graphene topological. 
\begin{figure*}[ht] 
  \centering
\includegraphics[width=4.83in]{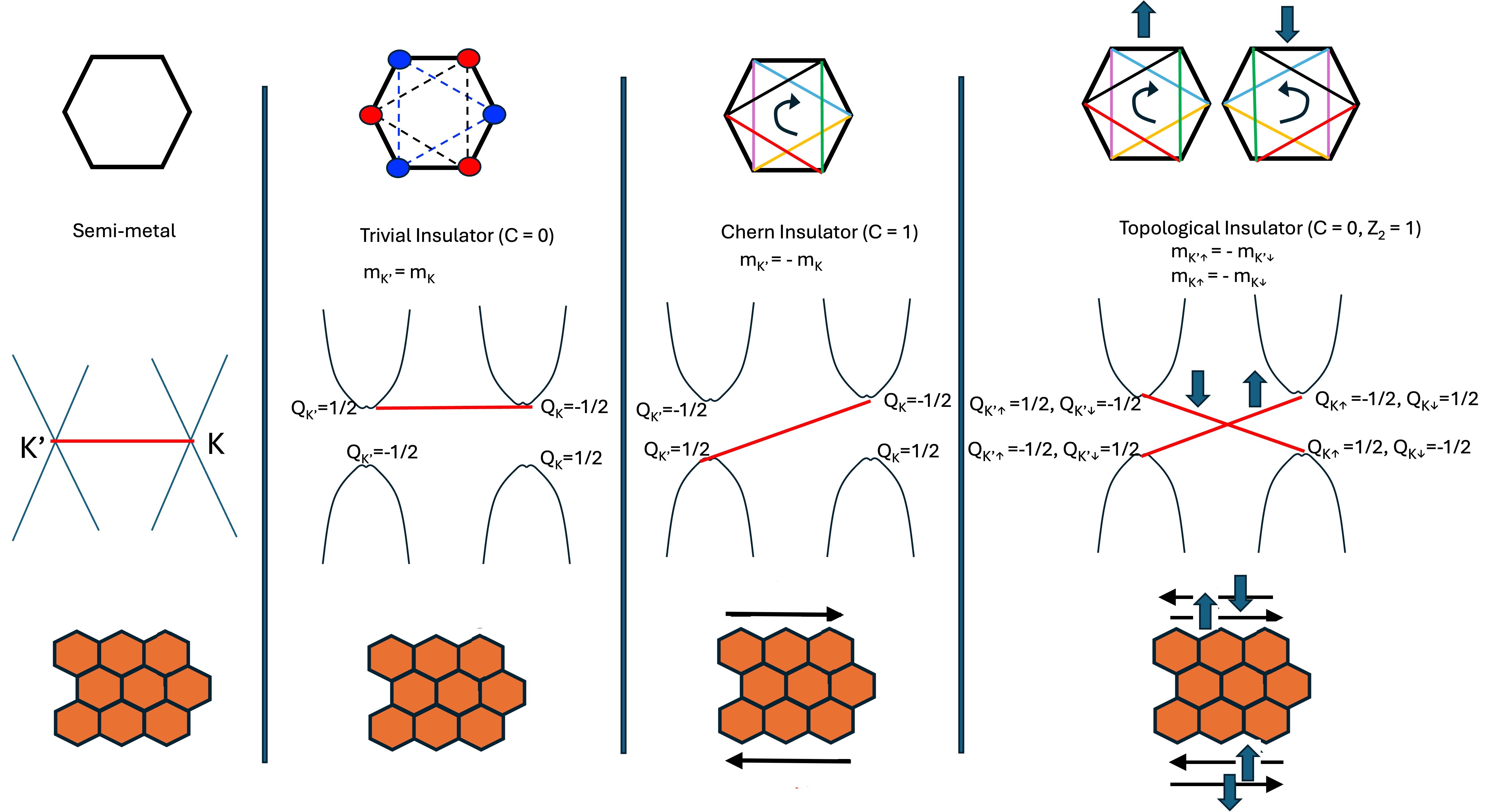} 
 \caption{Starting from gapless graphene with linear Dirac dispersion, successive perturbations modify the Dirac Hamiltonian to produce distinct topological phases. Sublattice symmetry breaking opens a trivial gap, while spin-orbit coupling generates a quantum spin Hall (topological insulator) phase with helical edge states. Time-reversal or inversion symmetry breaking splits the Dirac point into Weyl nodes of opposite chirality, producing Weyl semimetals characterized by Berry-curvature monopoles and topologically protected Fermi-arc surface states. The Chern number changes only when the bulk gap closes, illustrating how Berry curvature, symmetry, and topology determine the electronic phases.}
  \label{fig:top}
\end{figure*}
If we extend this picture to Bernal stacked bilayer graphene, then the out-of-plane p$_z$ orbitals vertically offset get pushed away and the A-B overlapping orbitals create a low-energy Hamiltonian that looks like 
\begin{equation}
    H_0(k) = \dfrac{\hbar^2}{2m}\left(\begin{array}{cc}0 & k_-^2\\k_+^2 & 0\end{array}\right)
\end{equation}
where $k_\pm = k_x\pm ik_y$ and $m \approx 0.03~m_0$. The two component pseudospin for this quadratic dispersion is 
\begin{equation}
      u_k = \dfrac{1}{\sqrt{2}}\left(\begin{array}{c}1 \\ \pm e^{\displaystyle 2i\theta_k}\end{array}\right), ~~~\theta_k = \tan^{-1}{(k_y/k_x)}
\end{equation}
Notably, the pseudospin winds twice  around the Fermi circle, so that $\pm k$ states now have the same pseudospin, still flipping between valleys, and between CB and VB (Fig.~\ref{fig:wind}, second panel).
We can generalize this to rhombohedral ABC-stacked N-layer graphene, 
whose low-energy chiral doublet scales as
$E \propto \pm |k|^N$, its pseudospin winding $N$ times around the Fermi surface. 

\subsection{Plenty of room at the edge}
We cannot continuously transform an orange into a donut without first shrinking it to a point then starting over (we are ignoring atomic structure). Analogously, we cannot smoothly transition between two distinct topological states without closing an energy gap separating them. The topological vs. non-topological behavior manifests clearly in its edge-states, something called the {\it{bulk-boundary correspondence}}.
If we fabricate a heterojunction with distinct topological segments side by side (say one with a positive mass and one with a negative mass), then at the interface we will necessarily need to have a vanishing mass for continuity, and an interface state connecting the band-edges carrying opposite mass on each side to allow a continuous transition. For a free-standing topological material surrounded by a non-topological material (e.g. vacuum), this manifests as a chiral edge state. 

The gaplessness of graphene arises from certain symmetries inherent in its lattice structure, which endows its eigenstates with orthogonalities (e.g. the pseudospins), so that the off-diagonal term responsible for level-repulsion vanishes. 
Consequently, opening a gap amounts to destroying certain symmetries in graphene.
A symmetry operation O corresponds to a conserved quantity, which allows us to block diagonalize the Hamiltonian using that label since there is no mixing between two different labeled sectors. 
This implies that O must  commute with the Hamiltonian in order to be preserved, since $dO/dt = [O,H]/i\hbar = 0$. 

Let us look at a sequence of increasing complexity that each arises from breaking progressively more symmetries in a zigzag graphene nano-ribbon, looking specifically for topological properties through its bulk-boundary correspondence. 
Along a zigzag graphene nanoribbon, there is an edge state that has zero band velocity (Fig.~\ref{fig:top} panel 1), connecting the Dirac points at the two inequivalent K and K$^\prime$ valleys, arising from the inequivalence of the sites along the boundary (A-type atoms appear with different frequency as B-type atoms). The question we can pose at this stage is if we break a symmetry that opens a band-gap in graphene, which of the two band-edges at each valley will the edge state remain connected to. 

\subsubsection{1. Graphene with broken sublattice symmetry} 
Nearest-neighbor graphene has a chiral, or sublattice, symmetry $\{\sigma_z,H_0\} = 0$, 
which pairs energies at $E$ and $-E$. Exchange of the A and B sublattices is represented by $\sigma_x$, but at fixed momentum it must be accompanied by the corresponding spatial or valley operation. Time reversal exchanges the two valleys; for spinless graphene it may be represented as $ T = \tau_xK$ ($K$: complex conjugation), while including real spin gives $T=is_y\tau_xK$. A staggered sublattice potential $M\sigma_z$
 breaks the chiral symmetry and A/B equivalence, opening a trivial gap, effectively turning graphene into boron nitride. 
\begin{equation}
    H = H_0 + M\sigma_z = \vec{d}\cdot\vec{\sigma}
\end{equation}
with energies $E = \pm \sqrt{\hbar^2v_x^2k_x^2 + \hbar^2v_y^2k_y^2 + m_\tau^2}$, $E_G = 2|m_\tau|$, where the Dirac mass $m_\tau = M$ imposes separate onsite energies $\pm M$, their average set to zero here. The coefficients $d_x, d_y, d_z$
 are the projections of an effective pseudomagnetic field onto the three Pauli matrices. As momentum changes, this field traces a path on the Bloch sphere. The in-plane winding of $(d_x,d_y)$ defines the chirality of the Dirac cone
\begin{equation}
    \chi_\tau = {\rm{sign ~det}}\left(\begin{array}{cc}\partial_{k_x}d_x & \partial_{k_x}d_y \\
    \partial_{k_y}d_x & \partial_{k_y}d_y \end{array}\right) = {\rm{sign}}(v_xv_y) 
\end{equation}
 while the sign of $d_z$ (the Dirac mass) determines which hemisphere the Bloch vector covers. 
For the occupied lower band, the half-integer contribution from that Dirac cone is, up to the overall Berry-curvature sign convention, $C_\tau = -\chi_\tau {\rm{sign}}(m_\tau)/2$, i.e., Chern number is product of chirality and sign of Dirac mass. 
For sublattice symmetry breaking, mass $m_\tau = M$ is the same for each valley, while the helicity $\gamma$ from the Berry phase is opposite. The net Chern numbers  are as a result opposite at the two valleys so that the total Chern number summed over the occupied valence bands at the two valleys vanishes and the corresponding Hall conductivity is zero. Boron nitride is therefore a trivial insulator. The edge state at the zigzag boundaries stays connected to either both conduction  or both valence band edges, depending on the sign of $M$, connecting two points with opposite Chern numbers $C = \pm 1/2$ and same mass. It is dispersionless, with zero group velocity as a localized state between the two projected valleys (Fig.~\ref{fig:top} panel 2). 

\subsubsection{2. Broken time-reversal symmetry (Chern insulators and the Haldane model)}
Magnetic fields break time-reversal symmetry, and we will notice the difference if we ran a film of graphene backward (up spins turn into down etc). Haldane argued \cite{haldane1988} that we can break this symmetry without an external field, preserving the essence of the Peierls phase factor by making the second nearest neighbor coupling complex, picking up a phase factor $t_2e^{i\phi}$. The graphene tight-binding Hamiltonian in the presence of both complex $t_2$ and field $M$ at low energy is 
\begin{equation}
    H = \mp \hbar v_F(k_x\sigma_x\pm k_y\sigma_y) + (\underbrace{M\mp 3\sqrt{3}t_2\sin{\phi}}_{\displaystyle m_\tau})\sigma_z
\end{equation}
The mass term multiplying $\sigma_z$ switches sign at $M = 3\sqrt{3} t_2|\sin{\phi}|$. 
Because the Dirac masses of the two valleys have opposite signs, their half-integer contributions add to a nonzero total Chern number. Bulk-boundary correspondence then requires a chiral state at a physical boundary with a trivial insulator or vacuum. 
A zigzag nanoribbon edge state couples a valence band top in one valley with the conduction band bottom in the other (Fig.~\ref{fig:top} panel 3), 
connecting opposite Chern numbers $\pm 1/2$. The linear dispersion endows these chiral states with a net velocity, moving either clockwise or anticlockwise depending on the sign of $t_2$, breaking time-reversal symmetry and creating a Chern insulator.  

\begin{table*}[t]
\begin{tabular}{llll}
\hline
System & Order parameter & $\mathbf{d}$ field & Topological texture \\
\hline
Graphene &
Pseudospin &
$\mathbf{d}=(k_x,k_y,0)$ &
Equatorial winding \\

Massive graphene &
Pseudospin &
$\mathbf{d}=(k_x,k_y,m)$ &
Gapped Dirac cone \\

Weyl semimetal &
Spin/Pseudospin &
$\mathbf{d}=(k_x,k_y,k_z)$ &
Momentum-space monopole \\

Topological insulator &
Spin &
$\mathbf{d}(\mathbf{k})$ &
Helical surface winding \\

Magnetic skyrmion &
Magnetization &
$\mathbf{d}(\mathbf{r})$ &
Real-space sphere wrapping \\
\hline
\end{tabular}
\end{table*}
\subsubsection{3. Restored time-reversal symmetry (Topological insulators and the Kane-Mele model).}

We can restore time-reversal symmetry by putting two identical copies of opposite Chern states on top of each other. One way to do this is to bring in electron spin, and making the hopping spin-dependent, so that the two opposite spins have opposite Chern behaviors \cite{kanemele2005}. This will happen in presence of spin-orbit coupling $\vec{L}\cdot\vec{S}$, which means $t_2$ flips sign for the two spin species. We will now have a clockwise spin-up edge state and anticlockwise spin-down edge state, restoring time-reversal symmetry while maintaining the gapped bulk state and helical edge states - counterpropagating Kramers partners with opposite spin. This is a topological insulator. The two spin valley components have opposite Chern number, so the net Chern number $C_\uparrow + C_\downarrow = 0$, but the states are helical and have a spin Chern number called the Z$_2$ index given by $Z_2 = (C_\uparrow - C_\downarrow)/2 = 1$ in the $s_z$-conserving limit. 
Outside that limit the individual spin Chern numbers need not be defined, whereas the $Z_2$
invariant survives under time-reversal symmetry. 
In effect, Chern number counts the number of twists in the wavefunction, while  $Z_2$ (integer $Z$ modulo 2) tracks whether it is odd or even. The spin states are now locked with momentum, up spins proceeding in a direction opposite to down spins (Fig.~\ref{fig:top} panel 4) in order to preserve time-reversal symmetry, a property called {\it{spin-momentum locking}}.

A 2-D HgTe layer has a negative mass since its bands are 'flipped' by strong spin-orbit coupling - the conduction band is p-type (odd parity) while the valence band is s-type (even parity) in orbital symmetry. Placing it next to a positive mass material like CdTe in a multi-quantum well system will move the valence band edge within a well above the conduction band edge. If we now expand the well size above a critical distance $d_c \sim 6.3$ nm, then the highest quantum confined hole state rises above the lowest electron state, inverting the discrete energy levels and creating a topological insulator with helical edge states, counterpropagating Kramers partners with opposite spin expectation values around the 1-D periphery. 
For a 3-D TI like Bi$_2$Se$_3$, there is a similar band-inversion. The bonding split $P^+$ orbitals of Bi and antibonding $P^-$ orbitals of Se are crystal field separated due to their planar stacking, isolating the $z$ orbitals. Spin orbit coupling inverts the levels, so the CB and VB have inverted parity within the $p_z$ manifold. 
As a result, they create 2-D opposite surface states at the top and bottom of the Bi$_2$Se$_3$ stack (Fig.~\ref{fig:wind}, panel 3).  

\subsubsection{4. 3D Dirac and Weyl semimetals }
The pseudospin of pristine graphene winds around a circle in the equatorial plane of the Bloch sphere with a Berry phase $\pi$, lacking any $\sigma_z$ term. Weyl semimetals reveal what happens when that winding escapes the plane and wraps the entire Bloch sphere: the Dirac point acquires the topology of a momentum-space magnetic monopole.
A 3-D Hamiltonian
\begin{equation}
    H = \hbar v_F(\sigma_xk_x + \sigma_yk_y + \sigma_zk_z)
\end{equation}
frees up the electron to explore the entire 3-D Bloch sphere. For graphene, the Berry curvature resembles the vector potential surrounding a solenoid. But for a Weyl node (the 3D equivalent of a Dirac point), it resembles the electric field of a magnetic monopole
\begin{equation}
    \vec{B}_k = \pm \vec{k}/2k^3
\end{equation}
Each sphere enclosing the node intercepts one quantum of Berry flux, with
$C = \pm 1$ denoting the monopole charge. 
A Weyl point is a twofold band crossing with definite chirality $C = \pm 1$. A Dirac point is a fourfold crossing that may be viewed as two opposite-chirality Weyl points superposed at the same energy and momentum. 
The double degeneracy means H must act on $\tau\otimes\sigma$ \cite{armitage2018weyl}
\begin{equation}
    H_D = \hbar v_F\tau_x\otimes (\vec{\sigma}\cdot\vec{k}) = \hbar v_F\left(\begin{array}{cc}0 & \vec{\sigma}\cdot\vec{k} \\\vec{\sigma}\cdot\vec{k}& 0\end{array}\right)
\end{equation}
Symmetry-breaking terms like $m\tau_z\otimes I + bI\otimes \sigma_z$ lift the coincidence of the two Weyl sectors. Depending on the representation and microscopic symmetry, the nodes may separate in momentum, shift in energy, or gap out. The energy eigenvalues of the perturbed Hamiltonian 
\begin{equation}
    \epsilon_{s,\mu}(\vec{k}) = s\sqrt{m^2 + b^2 + v_F^2p^2 + 2\mu b\sqrt{m^2 + v_F^2p_z^2}}
\end{equation}
Here $\vec{p} = \hbar\vec{k}$ while $s, \mu = \pm 1$ define the four branches. For $m > |b|$ we get a gapped semiconductor, while for $|b| > m$, we get two sets of Weyl points touching at $v_F\vec{p} = (0, 0, \pm \sqrt{b^2-m^2})$. These are simply the lowest-order symmetry-allowed couplings that distinguish the two chiral sectors. 

Let us visit the bulk-boundary correspondence with edge states at the surfaces of a stacked Weyl semimetal with the Hamiltonian $H_D + bI \otimes \sigma_z$. Imagine slicing momentum space. We specifically consider the Hamiltonian for a minimal time-reversal-broken Weyl semimetal
\begin{equation}
    H = k_x\sigma_x + k_y\sigma_y + [M(k_z) - B(k_x^2+k_y^2)]\sigma_z
\end{equation}
that can promote a gap closure. 
Each fixed $k_z \neq 0$ is a massive 2D Dirac model. Assuming $M(k_z) = k_0^2-k_z^2$, the gap closes with $M(k_z) = 0$, giving us two Weyl nodes at $k_z = \pm k_0$. 
The only thing that matters now is where the Bloch vector points at the center and at infinity.
If they point in opposite directions, the sphere must be wrapped once, while if they point the same direction, it is not wrapped. Since $M(k_z) = k_0^2 - k_z^2$,  the wrapping changes exactly at the Weyl nodes.
 The unit vector $\hat{d}$ goes from ${\rm{sign}}(M)\hat{z}$ near the origin to $-{\rm{sign}}(B)\hat{z}$ near $k_\perp \rightarrow \infty$, making the Chern number jump there. Indeed for each of those 2D slides in the $(k_x, k_y)$ plane for a fixed $k_z$, the Chern number $C(k_z) = 1/2[{\rm{sign}}(M(k_z)) + {\rm{sign}}(B)]$ up to an overall sign convention. For $B > 0$, we then get 
\begin{equation}
    C(k_z) = \begin{cases}1, ~~|k_z| < k_0 \\
    0, ~~ |k_z| > k_0\end{cases}
\end{equation}
Thus between the two Weyl nodes, each $k_z$	
-slice is a Chern insulator, while outside, it is trivial. Each $C=1$ 2-D slice has one chiral edge mode, so that stacking all the $k_z$ points between $\pm k_0$ at fixed energy creates a Fermi arc, terminated at the projected Weyl nodes, projecting to the opposite surface for an oppositely oriented arc, with spin-momentum locking (Fig.~\ref{fig:wind} top right panel). 

The two-node model is the minimal topology model for Chern slices and Fermi arcs. A TaAs-like inversion-broken, time-reversal-preserving Weyl semimetal is obtained by adding the time-reversed partner of that block, producing four Weyl nodes. Thus the Fermi-arc logic is unchanged, but the node multiplicity doubles.

\section{Applications: What does Topology buy an engineer?}
We now go beyond the tutorial stage of the paper to ask a pertinent question - how does the winding of the electron wavefunction impact the ability of a quantum material to accomplish something truly unique? It is often stated that topological ``protection'' protects the mobility and thus the ON current. 
ON current, however, is seldom the device bottleneck (electron speeds are orders of magnitude higher than tolerable gate frequencies). We argue that the novelty lies in the imposition of an added topological index that can control the OFF current across gate tunable barriers. These indices also allow electrons to selectively couple to specific symmetries (spins, polarizations), generating unique detectors and novel actuation mechanisms.

At a fundamental level topology can help with at least four mainstream attributes, each with a concrete application that we will present. The common  is that {\it{ there is a geometric constraint on underlying current flow imposed by the fact that wavefunction winding patterns need to match up smoothly around a tunable barrier, i.e., a topological index wants to be preserved.}}

The four examples below are organized by the physical quantity constrained by the underlying texture and the corresponding device function:
\begin{table}[h!]
\begin{tabular}{ll}
\hline
Topology Constrains... & Device Consequence \\
\hline
Deformation &
Memory\\
Transmission &
Switching \\
Symmetry Response & Actuation\\
Optical Selection Rules & Sensing\\
\hline
\end{tabular}
\end{table}

\subsection{Topology protects an isolated skyrmion from disappearing} 

\subsubsection{(i) Real-space winding and stabilization by the Dzyaloshinskii-Moriya interaction (DMI)}
As far as two-state binary logic or memory, the up and down magnetizations in a uniaxial anisotropic magnet can act as convenient stand-ins for electron charge. The anisotropy barrier separating the states is proportional to volume, so that an ultrascaled uniaxial magnet below $\sim 20$ nm tends to be highly fragile in its ability to hold a magnetization against thermal fluctuations, entering instead the `superparamagnetic' limit. Applications such as probabilistic and stochastic computing tend to capitalize on this volatility, where the states can act as `fair coins' that are truly random. However, for conventional memory applications, we need a way to restore the barrier for a small magnetic bit.

A magnetic skyrmion accomplishes just that property through its topological winding in 2-D {\it{real-space}} (the relevant SU(2) symmetry is that of its local spin texture) \cite{nagaosa2013topological}.  A skyrmion generates an emergent real-space Berry curvature whose integrated flux is fixed by a skyrmion number $N_{sk}$. As an electron moves through the texture, its spinor is parallel-transported over the Bloch sphere, and the accumulated Berry phase counts how many times the magnetization wraps that sphere. For a spinor corresponding to the magnetization unit vector
\begin{equation}
    \hat{m} = (\sin{\theta}\cos{\Psi},\sin{\theta}\sin{\Psi},\cos{\theta}), ~~~ u = \left(\begin{array}{c}\cos{\theta/2} \\ e^{i\Psi}\sin{\theta/2}\end{array}\right)
\end{equation}
the real-space Berry connection $\vec{A}(\vec{r}) = iu^*\nabla_ru = -(1-\cos{\theta})\vec{\nabla}\Psi/2$ up to an overall sign/gauge convention. The Berry curvature $B_z = \partial_xA_y - \partial_yA_x = -\sin{\theta}/2(\partial_x\theta\partial_y\Psi - \partial_x\Psi\partial_y\theta) = -\hat{m}\cdot(\partial_x\hat{m} \times \partial_y\hat{m})$/2. The skyrmion number
\begin{equation}
    N_{sk} = \int dxdy~~ \hat{m}\cdot\left({\partial\hat{m}}/{\partial x}\times {\partial\hat{m}}/{\partial y}\right)/4\pi
\end{equation}
At location $(r,\phi)$, the magnetization orientation is given by $(\theta,\Psi)$. A skyrmion's vortex-like winding motion makes the azimuth independent of angular orientation, $\theta = \theta(r)$, while the  in-plane tilt angle $\Psi$ should
increase linearly around the circle until it covers an integer multiple of $2\pi$, so that $\Psi  = n\phi + \psi$, with vorticity $n$ an integer and $\psi$ the domain angle or helicity (equal to $0$ and $\pi$ for N\'eel and $\pm \pi/2$ for Bloch skyrmions). A few steps of algebra then gives us $N_{sk} = [m_z(r=\infty)-m_z(r=0)]\times n/2$. So for a skyrmion whose core orientation is inverted relative to the background (an island of ups in a sea of downs), we get $N_{sk} = 1$. 
\begin{figure}[!h] 
  \centering
\includegraphics[width=3.2in]{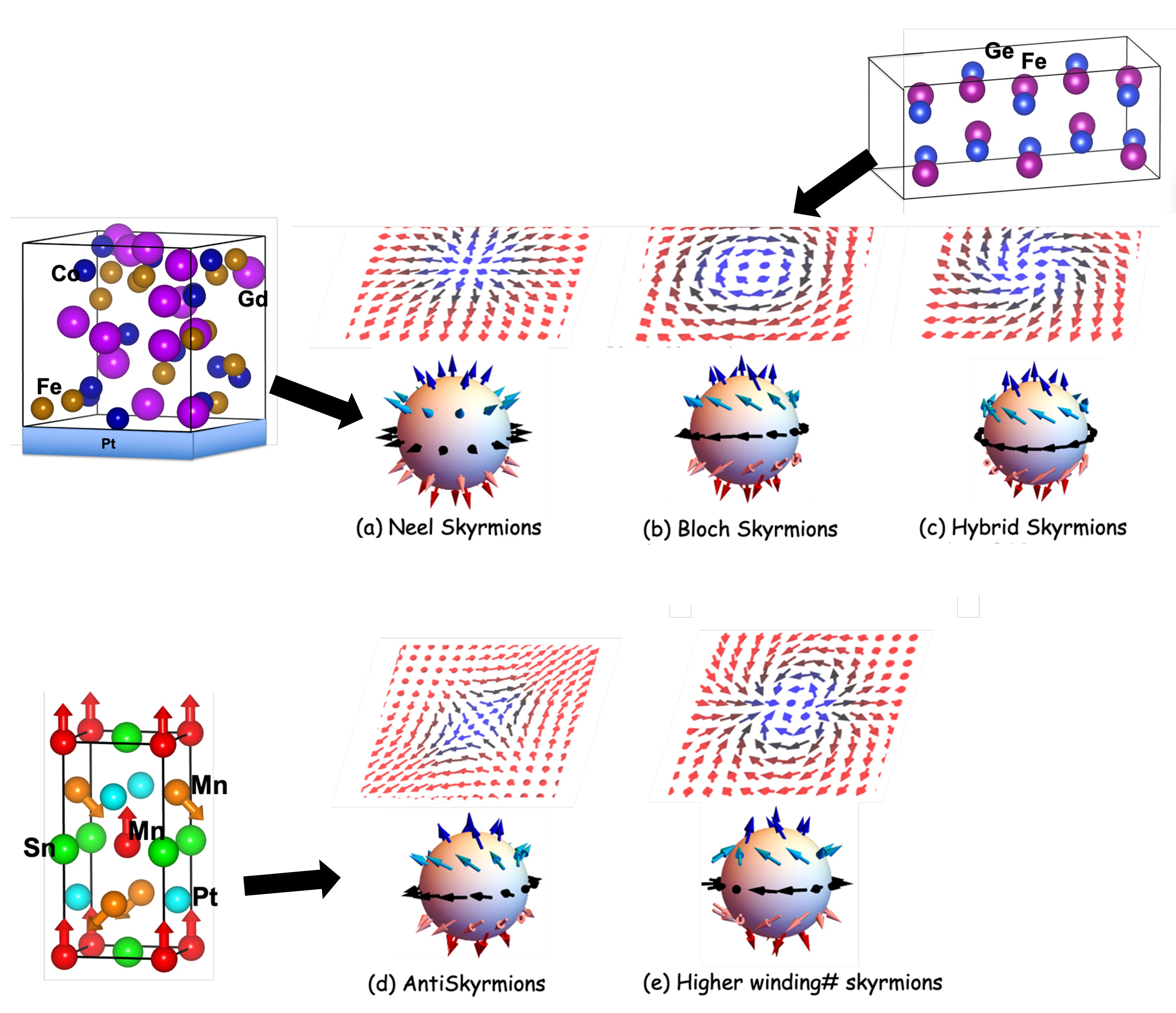} 
 \caption{The symmetry of the underlying crystal or interface fixes the orientation of the DMI vector D, which in turn dictates the energetically preferred sense of spin rotation. Interfacial inversion-symmetry breaking in heavy-metal/ferrimagnet heterostructures such as Pt/CoGd stabilizes Néel skyrmions, where spins rotate in radial planes. Bulk chiral B20 compounds such as MnSi and FeGe favor Bloch skyrmions with tangential spin rotation. Tetragonal $D_{2d}$
 Heusler compounds such as Mn
2PtSn stabilize antiskyrmions through anisotropic DMI, while intermediate combinations of interfacial and bulk DMI generate hybrid skyrmions with continuously varying helicity. Higher-winding textures ($|N_{sk}| > 1$) are also possible, although they frequently relax into lower-energy collections of unit-charge skyrmions. Thus, the microscopic symmetry of the material determines the real-space Berry-phase winding and the resulting topological texture.}
  \label{fig:skm}
\end{figure}

The driving field for the skyrmion's topological texture is a symmetry-breaking Dzyaloshinskii-Moriya interaction $E_{DMI} = \vec{D}\cdot(\vec{S}_i\times\vec{S}_j)$. The DMI can be interfacial as for CoGd on a heavy metal like Pt with high spin-orbit-coupling that generates isolated N\'eel skyrmions. It can also be bulk inversion-symmetry breaking such as B20 solids like FeGe or MnSi that creates stable Bloch skyrmions, usually lattices. Finally, D$_{2d}$ tetragonal unit cells like Mn$_2$PtSn have natural asymmetries that support antiskyrmions. In short then \cite{vakili2021skyrmionics}
\begin{equation}
    \vec{D} = \begin{cases}
    \hat{z}\times\vec{r}_{ij} = \pm \left(\begin{array}{c}-y \\ x\end{array}\right), ~~~\text{Interfacial, N\'eel}\\
    \vec{r}_{ij} = \pm \left(\begin{array}{c}x \\ y\end{array}\right), ~~~{\rm{(B20, Bloch)}}\\
    \sigma_z\cdot\vec{r}_{ij} = \pm \left(\begin{array}{c}x \\ -y\end{array}\right), ~~~(D_{2d}~{\rm{Antiskyrmions)}}
    \end{cases}
\end{equation}
To get the lowest energy then, the favored spin-rotation plane is perpendicular to $D$. To visualize the in-plane winding, we introduce an auxiliary planar field $\vec{r}=(m_x,m_y)$ and parameterize its integral curves by $s$, $d\vec{r}/ds = A\vec{r}$ to get a phase-portrait representation of the magnetization texture, giving us \cite{vakili2021skyrmionics}
\begin{equation}
    A = \begin{cases}
    \pm ~I ~~{\text{N\'eel}}\\
    \pm i\sigma_y~~{\rm{Bloch}}\\
    \pm \sigma_x ~~{\rm{Antiskyrmion}}\\
    \end{cases}
\end{equation}
The 2-D phase portraits of the winding patterns are then set by the eigenvalues of $A$, which in turn are uniquely determined by its trace and determinant. 
The A matrices describe the 2-D $S^2$ projections, the so-called Poinc\'are map of the skyrmions (Fig.~\ref{fig:skm}). The diagonal entries control radial expansion and contraction while off-diagonals control rotation. For an orientation preserving matrix $A_+ = \left(\begin{array}{cc}a & -b\\ b & a\end{array}\right)$ with positive determinant, the solution to $d\vec{r}/ds = A\vec{r}$ in polar coordinates is $r(\phi) = r_0e^{a\phi/b}$. This gives us a ray-like Ne\'el texture $y = Cx$ for $b = 0$, a circle-like Bloch texture $x^2 + y^2 = C^2$ for $ a = 0$, and a spiral-like hybrid in between. For an orientation-reversing matrix such as $A_- = \left(\begin{array}{cc}a & b\\b & -a\end{array}\right)$ we get $b(x^2 - y^2) + 2axy = C$, a family of hyperbolas around a saddle point, i.e., anti-skyrmions that reverse orientation between two orthogonal directions. 

\subsubsection{(ii) Topology induced energy barrier}
The Euler-Lagrange equation follows from a domain wall energy per unit area with exchange and anisotropy $E^\prime = t_F\int dx[A_{ex}(d\theta/dx)^2-K_u\cos^2{\theta}]$. Minimizing this energy 
 gives us $d^2\theta/dx^2 = \sin{\theta}\cos{\theta}/\Delta_0^2$ with domain wall width $\Delta_0 = \sqrt{A_{ex}/K_u}.$ Solving this equation with boundary condition $d\theta/dx|_{x=\infty} = 0$ yields a 1-D domain wall of the form $\theta(x) = 2\tan^{-1}[e^{(x-x_0)/\Delta_0}]$. If we now generalize these expressions to 3D including DMI, Zeeman and demagnetization fields, the resulting skyrmion shape function can be roughly described as two opposing domain walls each of width $\Delta$ at antipodal locations separated by a distance $R_{sk}$, the so-called `$2\pi$ model'
\begin{equation}
    \theta(r) = 2\tan^{-1}\left[\dfrac{\sinh{(R_{sk}/\Delta)}}{\sinh{(r/\Delta)}}\right]
\end{equation}
Under this {\it{ansatz}} for large skyrmions, the exchange DMI, uniaxial anisotropy $K_u$, demagnetization and Zeeman field $H_z$ contributions simplify to give a total energy \cite{vakili2021skyrmionics}
\begin{eqnarray}
 E_{skm} &=& E_{ex} + E_{DMI} + E_{ani} + E_{demag} + E_{Zeeman} \nonumber\\
    E_{ex} &\approx& 2\pi A_{ex}t_F\left(\dfrac{2R_{sk}}{\Delta} + \dfrac{2\Delta}{R_{sk}}N_{sk}^2\right)\nonumber\\
    E_{DMI} &\approx & -(2\pi R_{sk}t_F)\pi DN_{sk}\nonumber\\
    E_{ani} &\approx & (4\pi K_ut_F)R_{sk}\Delta \nonumber\\
    E_{demag} &=& -(2\pi\mu_0M_0^2t_F)R_{sk}\Delta\nonumber\\
    E_{Zeeman} &\approx & (\pi R_{sk}^2t_F) \times 2\mu_0M_0H_z
    \label{eq:energ}
\end{eqnarray}
$M_0$ is the saturation magnetization and $t_F$ film thickness.

Most relevant to this article  is the observation that the two main terms generating a parabolic well for an isolated metastable N\'eel skyrmion are the DMI-induced negative line tension $\propto -\pi DN_{sk}$, and the 2-D topological centrifugal exchange penalty $\propto  N_{sk}^2\Delta/R_{sk}$ from the angular winding $\langle \sin^2{\theta}\rangle/r^2$. 
The dependence on $N_{sk}$ suggests that topology partitions the continuum configuration space into sectors, while exchange, DMI, anisotropy, dipolar fields and atomic-scale collapse paths set the finite activation barrier separating them.

\begin{figure*}[ht!] 
  \centering
\includegraphics[width=6.83in]{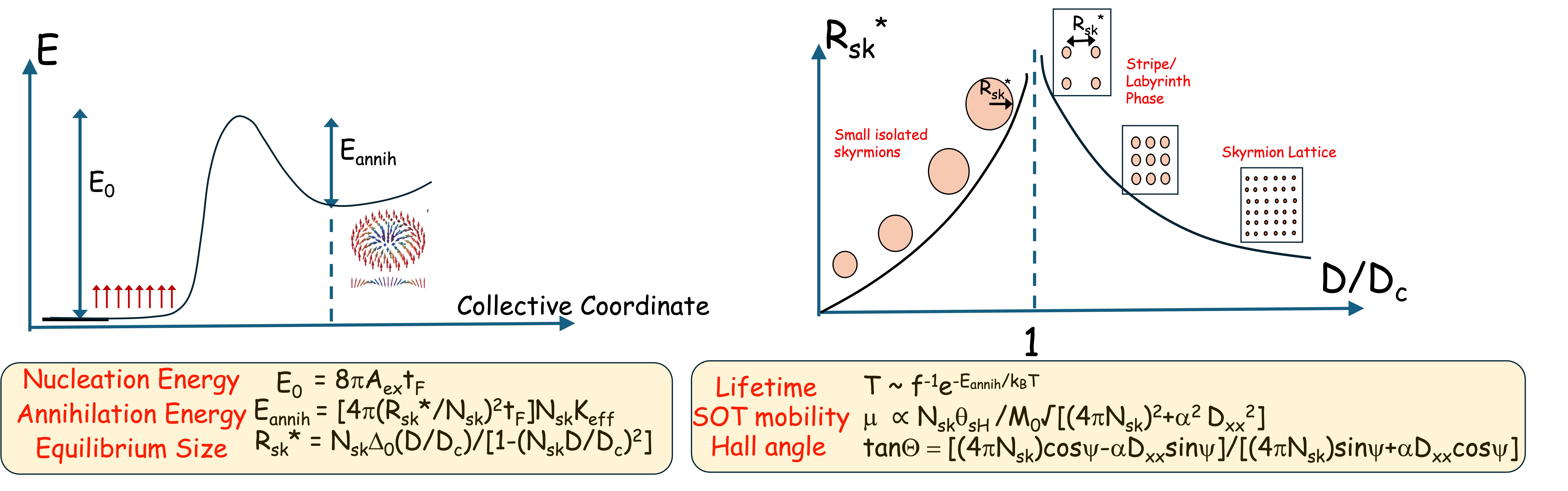} 
 \caption{Topology simultaneously determines every skyrmion metric including (a) the energy, (b) equilibrium size, thermal lifetime, and dynamical response of an isolated skyrmion through the winding number $N_{sk}$ \cite{vakili2021skyrmionics}.}
  \label{fig:ER}
\end{figure*}

We can now directly quantify the role of topology in endowing the skyrmion with an added stability barrier. The collapse saddle or Belavin–Polyakov energy scale is dominated by the exchange cost of concentrating a nonzero winding into a microscopic region
\begin{equation}
    E_{0} = A_{ex}t_F\int d^2\vec{r}(\nabla \vec{m})^2 \approx 8\pi A_{ex}t_F
\end{equation}The skyrmion minimum is lowered by DMI and shaped by anisotropy and demagnetization. At that minimizing skyrmion radius $R_{sk}^*$, we get approximately
\begin{equation}
    E_{skm} \approx E_{0}|N_{sk}|\sqrt{1-\left(\dfrac{|N_{sk}|D}{D_c}\right)^2}
\end{equation}
where $D_c = 4\sqrt{A_{ex}K_{eff}}/\pi$. 
This gives us the well depth or skyrmion annihilation barrier
\begin{equation}
    E_{annih} \approx  E_{0}|N_{sk}| - E_{skm}
\end{equation}
For $D \ll D_c$, the skyrmion annihilation barrier
\begin{equation}
    E_{annih} \approx \dfrac{\pi^3t_FD^2}{4K}|N_{sk}|^3
\end{equation}
\begin{figure*}[ht!] 
  \centering
\includegraphics[width=4.83in]{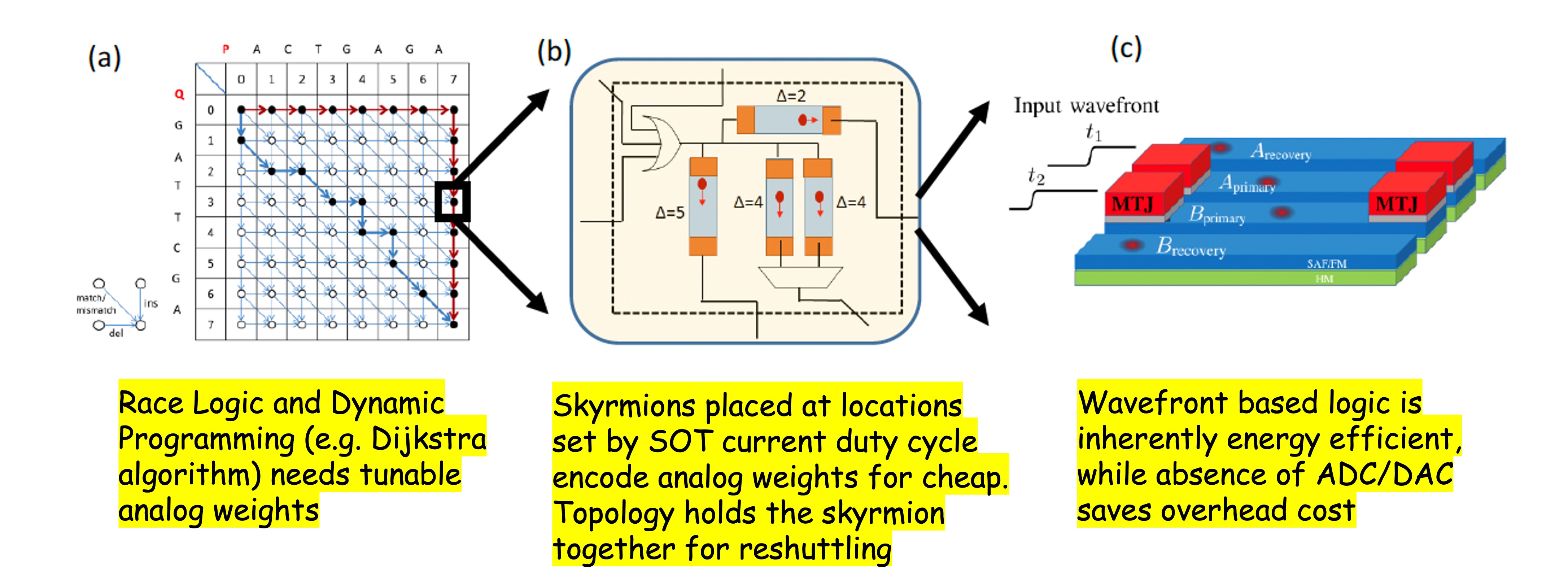} 
 \caption{Topology creates a metastable information state. Tunable analog weights required for race logic and dynamic programming are encoded in skyrmion position, set directly by the SOT-current duty cycle. Topology keeps each ultrasmall skyrmion intact during repeated reshuttling, while wavefront-based computation limits activity to the advancing front and the native analog representation avoids costly ADC/DAC conversion \cite{vakili2020temporal}}
  \label{fig:skm_mem}
\end{figure*}

\noindent Around the well bottom, minimizing $\partial E_{skm}/\partial R_{sk}$, we also get the skyrmion size
\begin{equation}
    R_{sk}^* = |N_{sk}|\Delta /\sqrt{1-(|N_{sk}|D/D_c)^2}
\end{equation} where $\Delta = |N_{sk}|(D/D_c)\Delta_0$, $\Delta_0 = \sqrt{A_{ex}/K_{eff}}$, with the effective anisotropy from Eq.~\ref{eq:energ} $K_{eff} \approx K_u - \mu_0M_0^2/2$. 
As $D$ approaches $D_c$, the well deepens until the skyrmion energy approaches that of the uniform state while its radius diverges, signaling the loss of positive domain-wall tension and the onset of extended modulated textures, whereupon $R_{sk}$ blows up and for larger $D$ creates a labyrinthine pattern of skyrmion droplets (Fig.~\ref{fig:ER}). 

Rewriting in terms of the optimized radius $R^*_{sk}$, 
\begin{eqnarray}
    &&E_{annih} = E_{0}|N_{sk}|\left[1-\sqrt{\dfrac{A_{ex}}{A_{ex}+K_{eff}(R^*_{sk}/|N_{sk}|)^2}}\right] ~~~~~~~\nonumber\\
    &&\overset{ R_{sk} \rightarrow 0}{\longrightarrow} \left[4\pi (R^*_{sk})^2t_F\right]K_{eff}/|N_{sk}| \propto |N_{sk}|
\end{eqnarray}
i.e., skyrmion volume times anisotropy. 
This higher-winding extension assumes a circular, unsplit $N_{sk}$-charge texture. For $|N_{sk}| > 1$, splitting into unit-charge skyrmions may provide a lower-energy pathway.

One challenge we see right away
is that  skyrmions with smaller $R^*_{sk}$ have shallower wells and shorter lifetimes. 
The way then to create a small, stable skyrmion with deeper wells is to increase the film thickness $t_F$ and also the effective anisotropy $K_{eff}$, by tunably reducing the demag term $\propto -M_0^2$. One way to do this is to approach a magnetization compensation point in a ferrimagnet like CoGd, suppressing the dipolar penalty and increasing $K_{eff}$. In CoGd, magnetization and angular-momentum compensation occur at distinct but compositionally nearby points; the latter is especially important for dynamics, while the former controls the demagnetization energy.
Indeed, the smallest skyrmions at room temperature, at around $\sim 10$ nm \cite{caretta2018fast}, have been seen in CoGd around magnetization compensation. Furthermore, for thick $\sim 15$ nm films $\sim 20$ nm skyrmion lifetimes have been estimated to be around 1 year for interfacial DMIs around 1 mJ/m$^2$. 

Compactifying the magnetic film by identifying the uniform boundary at infinity turns real space into a sphere $S_{real}^2$. The magnetization is therefore a map $S_{real}^2 \rightarrow S_{spin}^2$, whose continuous deformation classes form $\pi_2(S_2) = \mathbb{Z}$:
each texture is labeled by an integer $N_{sk}$, counting how many times real space wraps the magnetization sphere. A continuous deformation of the skyrmion simply rotates the spins on the sphere, preventing its annihilation. Destroying a skyrmion would either require shrinking it through a Bloch point to an atomistic limit that bypasses these continuum constraints (the gap closure described earlier), or collisions at high speed with edges and defects where finite size effects prevent full coverage of the Bloch sphere and hence the topological protection. At modest speeds, a skyrmion tends to deform its way around constrictions and holes, meaning it has fewer opportunities for pinning than the corresponding 1-D domain wall. Indeed, measured skyrmion mobility curves are fairly smooth. 

\subsubsection{(iii) Gyrotropic/Topological dynamics and damping}
The topological index also controls the longitudinal and transverse motion of the skyrmions driven by spin-orbit torque from current density $j_{HM}$ in an underlying heavy metal. We can start with the LLG equation \cite{ghosh2023}
\begin{eqnarray}
    \dfrac{d\hat{m}}{dt} = -\gamma_0\mu_0(\hat{m}\times\vec{H}_{eff}) + \alpha\hat{m}\times\dfrac{d\hat{m}}{dt} \nonumber\\
    - \dfrac{\hbar}{2q}\theta_{SH}\dfrac{\gamma}{M_0t_F}j_{HM}\left[\hat{m}\times(\hat{m}\times \vec{P}) + \eta \hat{m}\times\vec{P}\right]
\end{eqnarray}
where the effective Zeeman field $\vec{H}_{eff} = -(\delta E/\delta \vec{m})/\mu_0M_0t_F$, $\gamma_0$ is the gyromagnetic ratio, $\vec{P}$ is the electron spin polarization which we will assume to be along the $\vec{j}_{HM}\times\hat{z}$ direction, while the non-conserved parts are set by the Gilbert damping parameter $\alpha$ and the current density $j_{HM}$ in the heavy metal and the spin Hall angle $\theta_{SH}$ that sets the spin-orbit torque. $\eta$ is the non-adiabaticity parameter. 

In the rigid approximation the LLG equation reduces to the Thiele equation that describes how the SOT force rotates with helicity \cite{vakili2020self}
\begin{equation}
    \vec{G}\times \vec{v} - \alpha ({\cal{D}}.\vec{v}) + BR(\psi)\vec{j}_{HM} = 0
\end{equation}
with the gyrotropic vector $\vec{G} = (0,0,-4\pi N_{sk})$, tilt angle $\psi$ that generates a  2-D rotation matrix R, $\theta_{SH}$ is the spin Hall angle from the heavy metal set by its spin orbit coupling, and the dissipation tensor ${\cal{D}}_{ij} = \int dxdy\partial_i\vec{m}\cdot\partial_j\vec{m}$, which for our circular $2\pi$ model is roughly ${\cal{D}}_{xx} \propto R_{sk}^*/\Delta $. The SOT pre-factor $B = \pi^2 \theta_{SH}\hbar I_d\gamma_0/2q M_0t_{F}$, where $I_d = \int dr (r\partial_r\varphi + N_{sk}\sin{\varphi}\cos{\varphi}) \approx \pi N_{sk}R_{sk}$ set by the angle $\varphi$ between $\hat{m}$ and $\hat{z}$. 

The deflection by Magnus force is set by $N_{sk}$ through the gyrotropic vector $\vec{G}$, while the SOT force direction (last term in the Thi\'ele equation) is set by tilt angle $\psi$. The
resulting current density can be related to the skyrmion carrier density, $j_{HM} R(\psi-\theta_0)= qn_{sk}v_{sk}$, with a 2-D rotation matrix $R$ that depends on tilt angle $\psi$ and a deflection angle $\theta_0$ satisfying $\tan{\theta_0} = 4\pi N_{sk}/\alpha {\cal{D}}_{xx}$. The topologically relevant mobile charge density is given by angular momentum conservation, $n_{sk}(\hbar/2)\theta_{SH} = (M_0t_F/\gamma_0)\sqrt{(4\pi N_{sk})^2 + \alpha^2D_{xx}^2} $. This then sets the skyrmion velocity
\begin{equation}
    \vec{v} = \dfrac{\gamma_0\pi^2R_{sk}(\hbar/2q)\theta_{SH}}{M_0t_F\sqrt{(4\pi N_{sk})^2 + \alpha^2{\cal{D}}_{xx}^2}}N_{sk}R(\psi-\theta_0)\vec{j}_{HM}
\end{equation}The skyrmion 
can be sped up near magnetization compensation in ferrimagnets  where a suitable generalization replaces $M_0/\gamma_0$ with $M_1/\gamma_1 - M_2/\gamma_2$ for the two sublattices.
Near compensation, CoGd ferrimagnetic skyrmions have been driven above 600 m/s \cite{quessab2022zero}, while related ferrimagnetic CoGd and insulating-garnet domain walls have reached approximately 1 km/s \cite{avci2019interface}.
More significantly, the skyrmion is seen to
have an added topological slow-down given by the $N_{sk}$ term in the denominator, relative to 1-D domain walls. The equations however, need to be generalized to ferrimagnets by suitable averaging over composition, as described in \cite{vakili2021skyrmionics}.
\begin{figure*}[bt!] 
  \centering
\includegraphics[width=4.83in]{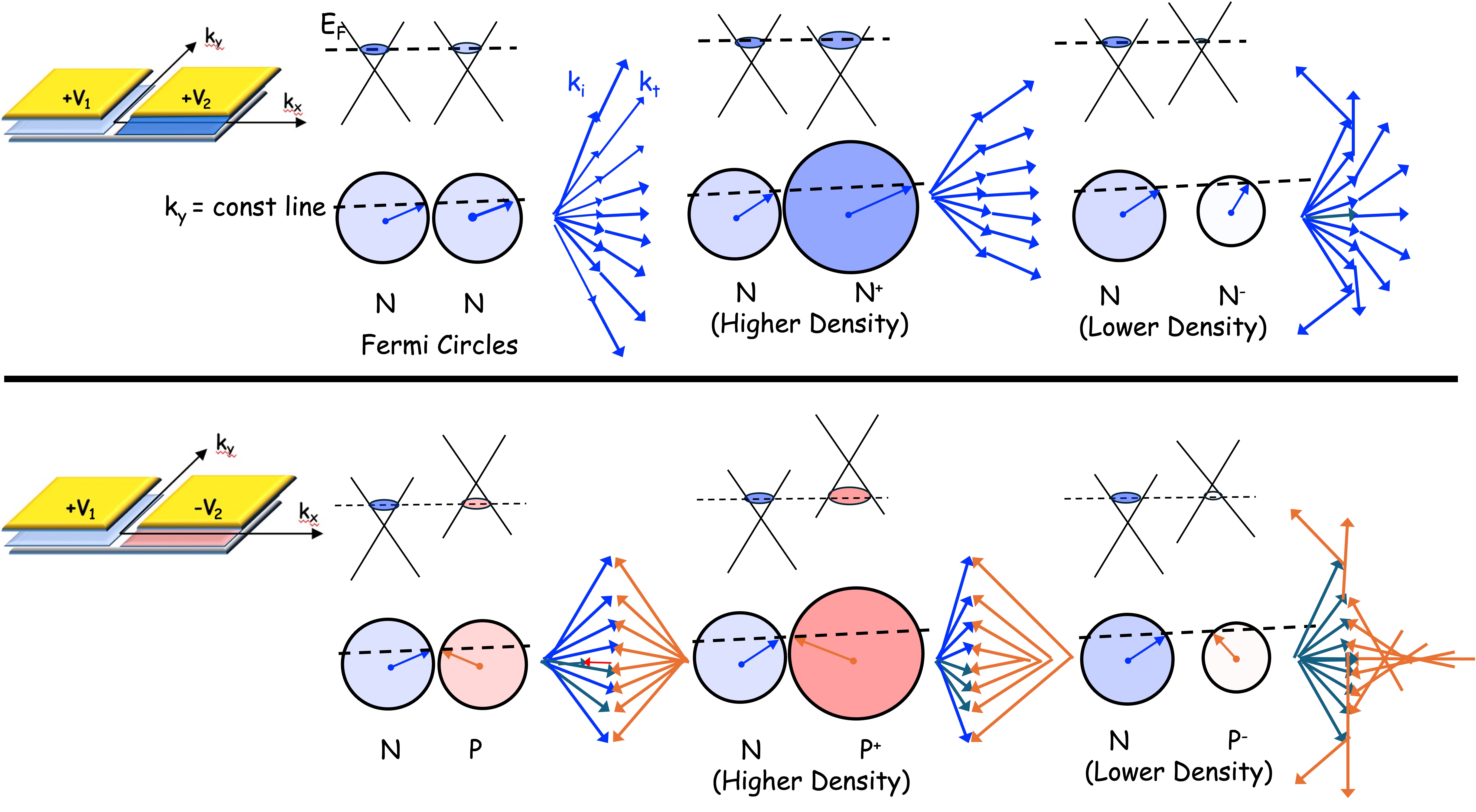} 
 \caption{Top: Positive-index (n-n) junctions. Bottom: Negative-index (n-p) junctions. The upper row in each panel shows the corresponding Dirac cones with the Fermi level $E_F$, while the lower row shows the associated constant-energy contours (Fermi circles) in momentum space. Blue and orange arrows denote the incident and transmitted crystal momenta, respectively. Conservation of the transverse wavevector, $k_y = $const, uniquely determines the transmitted momentum for each incident state. For n-type graphene, the group velocity is parallel to the crystal momentum, whereas for p-type graphene it is antiparallel because transport occurs in the valence band, giving rise to negative electronic refraction. Equal carrier densities ($k_{Fn} = k_{Fp}$)
produce the symmetric Veselago lens ($n_{eff} =-1$), while unequal densities (N$^\pm$ or P$^\pm$)
 modify the effective refractive index $n_{eff} = \pm k_2/k_1$, shifting the paraxial focal position and ultimately leading to total internal reflection as the transmitted carrier density approaches zero. Experimental signatures of this behavior exist in magnetoconductance measurements. }
  \label{fig:pnj}
\end{figure*}
The rotation matrix also gives us a transverse Magnus force that gives a skyrmion Hall angle $\Theta_{skH}$ that depends on the tilt (N\'eel vs Bloch)
\begin{equation}
    \dfrac{v_{sk,y}}{v_{sk,x}} = \tan{\Theta_{skH}} = \dfrac{4\pi N_{sk}\cos{\psi}-\alpha {\cal{D}}_{xx}\sin{\psi}}{4\pi N_{sk}\sin{\psi}+\alpha {\cal{D}}_{xx}\cos{\psi}}
\end{equation}
There is a particular tilt angle $\psi_c = \tan^{-1}(4\pi N_{sk}/\alpha {\cal{D}}_{xx})$ in a hybrid skyrmion where this angle vanishes and the Magnus force aligns with the drive direction of the current \cite{vakili2020self}. The tilt angle is set by the DMI energy density $\epsilon_{DMI} = D_{int}\cos{\psi}-D_{bulk}\sin{\psi}$ (all bulk for Bloch, $\psi = \pm \pi/2$, all interface for N\'eel, $\psi = 0, \pi$). Minimizing, we get the optimized tilt angle $\psi_0 = \tan^{-1}(-D_{bulk}/D_{int})$. 
We can make $\psi_0$ approach the compensation point $\psi_c$ by using a material with a graded DMI, such as $Pt_xW_{1-x}$ in the underlying heavy metal along the current direction, or a bulk graded composition $Fe_yCo_{1-y}$ along the growth direction. 

\subsubsection{(iv) Device application $\rightarrow$ native temporal memory} The tunable quasi-ballistic nature,  like beads on an abacus, allows a skyrmion to encode the duty cycle of an SOT current pulse in its location along a racetrack (Fig.~\ref{fig:skm_mem}).  A skyrmion can thus be used as a native temporal memory \cite{vakili2020temporal} for in-sensor data processing as well as MIN/MAX type dynamic programming, where the solutions of sub-computations are partial solutions of total computation (e.g. the Dijkstra algorithm). Since computation is only at the wavefronts, this saves energy, while the natural analog nature of skyrmions avoids costly analog-to-digital (ADC) or DAC components. 

Here topology serves to stabilize an ultrasmall, movable magnetic texture whose position can encode information. This enables native temporal and spatial memory for unary and wavefront-based computation, although nondestructive readout may require a copy-and-hold architecture.

\subsection{Topology can prevent carriers in graphene from reflecting or transmitting}
We now switch to another application of topology - based on the SU(2) symmetry of pseudospins in graphene. 
Graphene reproduces geometrical optics in four successive approximations: ray optics (Snell's law), wave optics (Fresnel equations), polarization optics (Malus' law), and finally topological optics, where wavefunction winding itself becomes the conserved quantity. The pseudo-spin degree of freedom introduced in graphene is not merely a topological curiosity; it enables wave-optical phenomena including Klein tunneling, negative refraction, collimation, Veselago lensing, and ultimately transistor concepts based entirely on coherent Dirac-fermion optics.
\subsubsection{(i) Ray Optics $\rightarrow$  Snell's Law and negative refraction}
We can follow usual equations of geometrical optics for graphene, whose Dirac bandstructure resembles photons (albeit with antiparticles - electrons and holes). A local electrostatic gate shifts the entire Dirac cone along the energy axis, preserving however the opening angle and the group velocity, unlike photons that slow down under an electrostatic potential in a dense optical material. 

Let us consider a split-gated graphene NN or PP junction in the x-y plane, with gate interface along the y-axis. Translational symmetry along the interface means the transverse wave vector $k_y$ is conserved, while $k_x$ changes value across the interface in order to fit within the different fermi circle radii, giving us the conventional Snell's law with angles of incidence, reflection and transmission designated as $\theta_1$, $\theta_1^\prime$ and $\theta_2$ respectively
\begin{equation}
    \theta_1 = \theta_1^\prime, ~~k_{F1}\sin{\theta_1} = k_{F2}\sin{\theta_2}, ~~~ k_{F1,2} = (qV_{G1,2})/\hbar v_F 
\end{equation}
with the local gate voltage acting like a refractive index. 
For two gates with identical polarity, electrons behave like optical rays (Fig.~\ref{fig:pnj}, top panel), bending toward the normal when moving from lower electron density region (lower positive gate bias) to higher, and away from the normal going from higher density to lower  leading to total internal reflection above a critical angle of incidence. 

The situation becomes more interesting when we have opposite gate polarities creating a PN junction (Fig.~\ref{fig:pnj}, bottom panel). Since electron states in one side have to match winding with hole states in the other (CB vs VB), the group velocity switches sign and is antiparallel to $k_x$ on the P side while staying parallel on the N side. This means we get in effect, a negative index of refraction
\begin{equation}
    k_{F1}\sin{\theta_1} = -k_{F2}\sin{\theta_2} 
\end{equation}
For an equally doped PN junction ($V_{G1} = -V_{G2}$), the incident electron rays from a common point of origin bend and focus to a mirror point on the other side, leading to a {\it{Veselago lens}} \cite{veselago1968,cheianov2007}. For unequal doping, we get a paraxial defocusing as wider incident rays focus further and further at the denser $P^+$ and closer and closer for rarer $P^-$ side eventually leading to a glancing emergent ray at critical angle. 

Negative index of electrons in graphene has been experimentally demonstrated \cite{chen2016} using magnetoconductance experiments in a four-probe PN junction set up. The resonant peaks indicate captured electrons on the refracted side, and their measured voltage polarity indicates whether they were captured by an upper or lower probe, directly indicating whether the emergent electron bent along a positive or negative index trajectory. 

\subsubsection{(ii) Wave Optics $\rightarrow$ Anomalous Fresnel's Law, Brewster angles, and Klein/anti-Klein tunneling}
The pseudospin structure for electrons in monolayer graphene (MLG) is quite telling, The forward and reverse momenta have orthogonal Bloch states, preventing an electron from back-scattering at normal incidence. The reflection coefficient is set by the overlap of the Bloch wavefunctions $|u_1^*u_2|^2 = \cos^2{(\theta_1-\theta_2)/2}$. We match the spinors across the boundary, keeping in mind that across the boundary $k_y$ stays the same but $k_x$ flips sign \cite{ghosh2016}
\begin{equation}
\psi = \begin{cases}\left(\begin{array}{c}1 \\ e^{i\theta_1}\end{array}\right)e^{i(k_xx+k_yy)} + r \left(\begin{array}{c}1 \\ -e^{-i\theta_1^\prime}\end{array}\right)e^{i(-k_x^\prime x+k_y^\prime y)} \\
t\left(\begin{array}{c}1 \\ e^{i\theta_2}\end{array}\right)e^{i(-k_x^{\prime\prime} x+k_y^{\prime\prime}y)}
\end{cases}
\end{equation}
The geometrical Snell's law comes from matching the phases at $x = 0$ for all $y$, which yields $k_y = k^\prime_y = k_y^{\prime\prime}$ (witness the consequence of a constant $k_y$ in Fig.~\ref{fig:pnj}). Matching the pseudospins, however, gives us the equivalent of Fresnel's equations for EM fields, where we match tangential components of magnetic field and normal components of electric displacement vector across the interface. This gives us the reflection and transmission coefficients 
\begin{eqnarray}
    T &=& |t|^2\dfrac{v_{x2}}{v_{x1}} = |t|^2\dfrac{\cos{\theta_2}}{\cos{\theta_1}} = \dfrac{2\cos{\theta_1}\cos{\theta_{2R}}}{\cosh{\theta_{2I}} + \cos{(\theta_1+\theta_{2R})}} \nonumber\\
    R &=& |r|^2 = 1 - T
\end{eqnarray}
where the subscripts $R$ and $I$ indicate real and imaginary parts, and the angles $\theta_{1,2}$ are related by Snell's laws.

We can check the various limits easily - for a homogenous PP or NN junction, $\theta_1 = \theta_2$ and $T = 1$. For a symmetrically doped PN junction with $\theta_1 = -\theta_2$, we get $T = \cos^2{\theta_1}$. For $\theta_1 > \theta_C$, the critical angle when moving from denser to rarer, where large angle incident electrons in the larger Fermi circle on the left cannot conserve their $k_y$ values with the smaller Fermi circle on the right, $\theta_{2R} = \pi/2$ and $\theta_{2I} > 0$, at which point $T = 0$. 
Most significantly, at $\theta_1 = 0$, normal incidence, the reflectivity vanishes, as the backscattering is symmetry prohibited by the orthogonal pseudospin states within a single valley. This universal transmission $T(\theta_1 = 0) = 1$ is called {\it{Klein tunneling}}, and has the consequence of pinning it to unity, even while the higher angles can reflect. At this point, increasing the gate voltages influence the Snell's law ratio, {\it{effectively collimating the electrons.}} The collimation becomes even more aggressive for finite gate split $d$, which adds an extra tunneling pre-factor 
\begin{equation}
    T_{split-gate} = Te^{-2\pi dk_\parallel |\sin{\theta_1}\sin{\theta_2}|}, ~~k_\parallel = \dfrac{|k_{F1}k_{F2}|}{|k_{F1}|+|k_{F2}|}
\end{equation}
Summing over incident angles, we get a mode-averaged electron transmission
\begin{equation}
    \bar{T} \approx \dfrac{1}{|E|}\sqrt{\dfrac{U_0\hbar v_F}{d}}, ~~U_0 = q\Delta V_G = \hbar v_F(|k_{F1}|+|k_{F2}|)
\end{equation}
\begin{figure*}[ht!] 
  \centering
\includegraphics[width=4.83in]{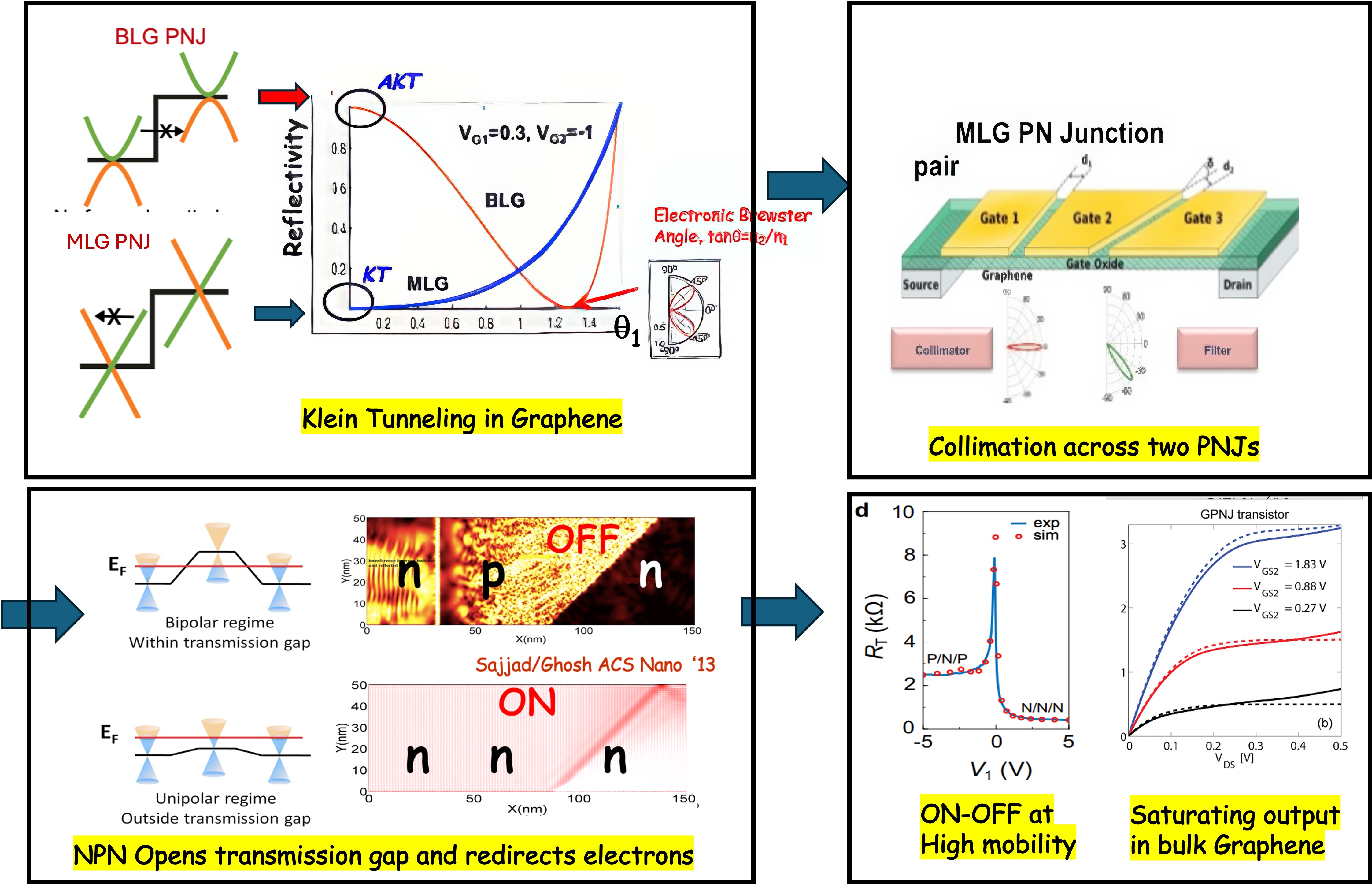} 
 \caption{Topology controlled transmission. (a) Klein Tunneling (KT) mono (MLG) as well as Anti-Klein Tunneling (AKT) and Brewster angles across  bilayer graphene (BLG) PN junctions impose respectively universal values of zero and unit reflectivity at normal incidence independent of barrier height (Fig.~\ref{fig:KTFET}(a)), driven by the matching of topological winding of the pseudospins across the junction. Experimental verification was done in Corbino disks. (b) The resulting collimation of higher angle electrons can be utilized in a transistor structure with two non-collinear split gate junctions where the collimation lobes do not align. (c) For nnn we get high transmission, while for npn the non-aligned collimation lobes cut off current by effectively opening a gate-tunable transmission gap (colorplots are atomistic quantum kinetic simulations of current density). (d) This can give an ON-OFF $\sim$ 10-13 at high mobility (theory and experiment), as well as an output resistance that is very high with a saturating $I-V$. Although the modest ON-OFF is not a game-changer in digital electronics, the modest ON-OFF and saturating I-V with high mobility yields a large power gain in an RF device, with the added bonus of integrability \cite{tan2017scirep,sajjad2013acsnano,elahi2024}.}
 \label{fig:KTFET}
\end{figure*}
Let us now repeat this analysis for bilayer graphene (BLG), whose pseudospins wind twice as fast. Going through the same exercise, for an abrupt, unbiased bilayer p-n step in the ideal two-band, isotropic approximation, the transmission takes the form 
\begin{equation}
    T_{BLG} = \dfrac{2\sin{2\theta_1}\sin{2\theta_{2R}}}{\cosh{2\theta_{2I}}+ \cos{(2\theta_1+2\theta_{2R})}}
\end{equation}
\begin{figure*}[ht!] 
  \centering
\includegraphics[width=4.83in]{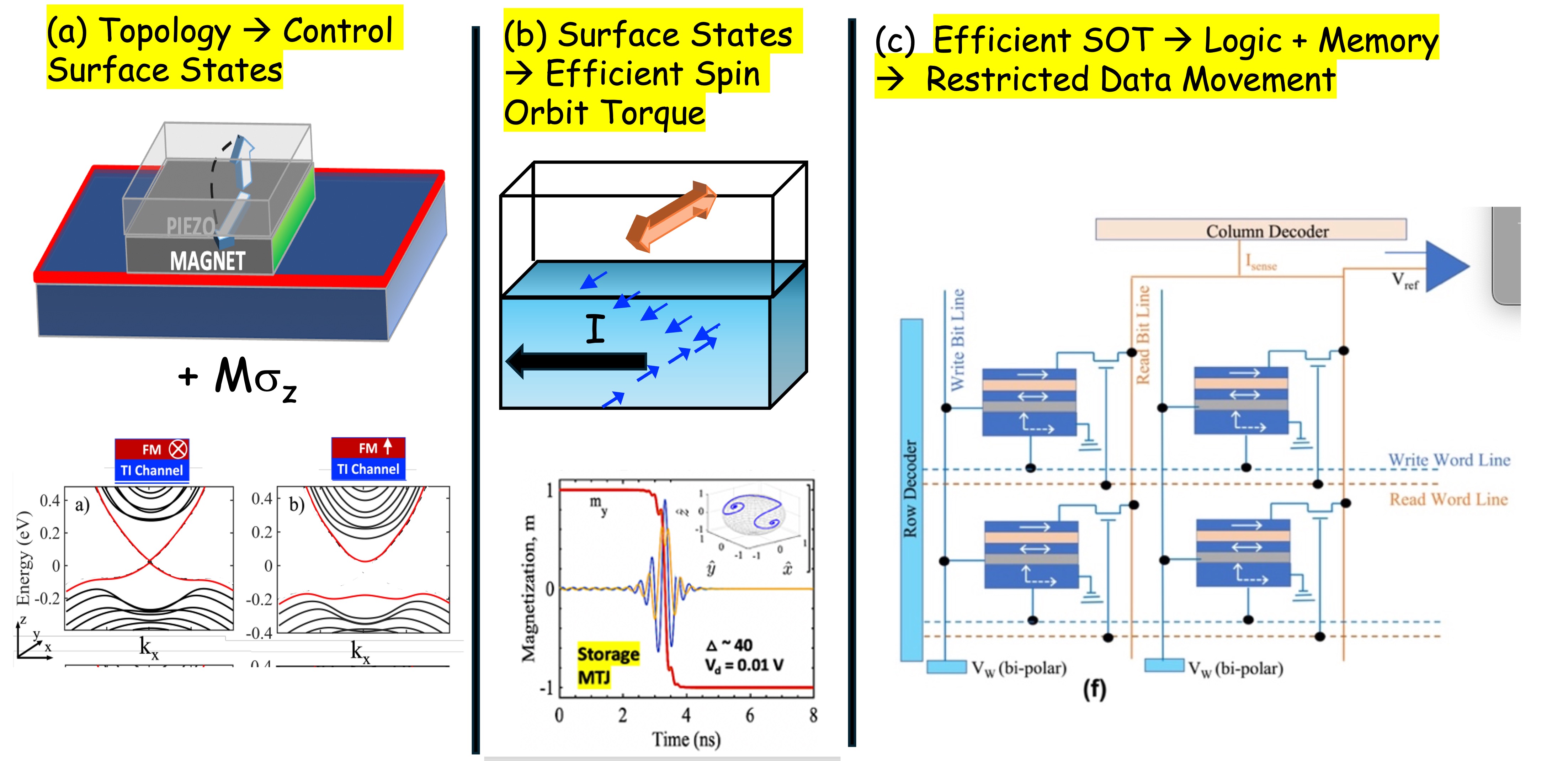} 
 \caption{Topology controlled actuation. Breaking or restoring the topological protection of topological-insulator and Weyl-semimetal surface states allows electrical gating of spin transport. The resulting spin-momentum locking converts charge current into deterministic spin accumulation and spin-orbit torque, while symmetry breaking in Weyl semimetals enables bulk spin Hall currents. Topology thus becomes a gate-controlled actuator, selectively generating the spin symmetries required for efficient processor-in-memory switching \cite{morshed2025prapplied}.}
 \label{fig:TImag}
\end{figure*}

\noindent For this system, normal incidence $\theta_1 = 0$ gives zero transmission and perfect reflection, known as {\it{anti-Klein tunneling}}. We can see this from Fig.~\ref{fig:wind} panel 2. Electron pseudospins are perfectly aligned at forward and reverse within a branch, but for a PN junction, the two branches being orthogonal we get perfect reflection but no transmission. This implies there is an intermediate sweet spot for transmission. Indeed, the maximum transmission happens at $\theta_B = \tan^{-1}{|V_{G2}/V_{G1}|}$. The physics here is the optical analogue of Brewster angles for perpendicularly polarized photons. In the electronic analogue, the reflection is quenched because the pseudospins of the incident and reflected electrons become orthogonal at 90$^0$ (Fig.~\ref{fig:wind} panel 2), when the reflection and transmission directions become orthogonal, so that $\theta_1 + \theta_2 = \pi/2$. These results have been confirmed experimentally \cite{elahi2024} with `edge-less' Corbino disks, whose magnetoconductance traces show respectively a fixed peak and a dip for MLG and BLG respectively, as predicted. 

If we can extend the ABC rhombohedral stacking to N-layers (ignoring correlation effects as in atomic hexagonal lattices), the PN transmission oscillates between the Klein/anti-Klein limits with increasing number of Brewster angles until the odd-N layer transmission approaches a cosine function and the even-N a sine function. 
\subsubsection{(iii) Polarization Optics $\rightarrow$ Anomalous Malus' Law}
What we just saw is that conservation of pseudospin in monolayer graphene pins the low-angle transmission and thus collimates electrons perpendicular to a split-gated PN junction with progressively increasing voltage gradient. 
Relative to electron incidence angle defined by a contact, if we rotate the junction by an angle $\delta$, then the mode summed conductance is given by \cite{sajjad2012prb}
\begin{equation}
    G \approx G_0 \int_{-\pi/2}^{\pi/2-\delta}\dfrac{T(\theta+\delta)}{\Delta \theta}d\theta = \dfrac{2M}{3}\cos^4{\left(\dfrac{\delta}{2}\right)}
\end{equation}
where $M$ is the mode count, $G_0 = 4q^2/h$ is the quantized conductance including valley and spin degeneracies, and the quantized angle $\Delta\theta = \Delta k_y/k_F\cos{\theta}$. This equation is the equivalent of Malus' law for a polarizer-analyzer, except we have half-angles involved, consistent with the SU(2) spinors. Experiments show this trend - a drop in transmission with angular orientation, with an intermediate rise from a parallel conductance channel from edge scattering that redirects electrons towards the junction. 

\subsubsection{(iv) Device Application $\rightarrow$ Klein tunnel transistor}
Graphene has a high mobility due to its low-energy massless Dirac bandstructure, at the expense of gaplessness. Opening a gap $E_G$ endows it with mass $E_G/2v_F^2$ resembling the rest energy of photons and the Higgs mechanism, robbing it of its high mobility. However, the topology of pseudospins and Klein tunneling offers a potential way past this bandgap-mobility trade-off \cite{sajjad2011apl,sajjad2013acsnano,wang2019}. Consider the wedge shaped gated structure in Fig.~\ref{fig:KTFET}(b). For uniform gating with positive voltage creating an $nnn$ structure, we expect high mobility. However by reversing the voltage of the central wedge, we set into motion Klein tunneling across the two PN junctions, and the non-alignment of their collimated transmission lobes shown in green and red redirects the electrons towards the source, in effect opening a gate tunable transmission gap of magnitude $E_G \approx \Delta V_G\sin^2{\delta}$ for an abrupt junction.  The resulting $I-V_G$ shows an ON-OFF $\sim 6-13$ (Fig.~\ref{fig:KTFET}(c), theory and experiment) in bulk graphene while maintaining a scattering length above 1 $\mu$m, while the computed output characteristic $I-V_D$ shows a saturation that is uncharacteristic of normal graphene. While the ON-OFF is modest (multiple reflections turn the gap into a pseudogap), Klein tunneling increases transconductance and creates a high output resistance, which together with high mobility potentially place the graphene Klein tunnel transistor in competition with other III-V based RF devices \cite{tan2017scirep}, with the added advantage of integrability with silicon. 

Unlike conventional bandgap engineering, Klein-tunneling transistors suppress current by increasing the intrinsic channel resistance through transmission filtering rather than carrier depletion. Consequently, once the OFF-state channel resistance exceeds the contact resistance, the latter ceases to dominate the device characteristics, relaxing one of the principal constraints that has historically limited many two-dimensional semiconductor technologies.

The role of topology here is not to create a band gap, but a transmission gap: pseudospin matching restricts the trajectories that can cross the junction. An otherwise metallic material can therefore be geometrically switched while retaining its massless Dirac dispersion.

\subsection{Topology can allow us to symmetry-gate a metallic state}

We have already seen how we can open a trivial gap in the surface states of a topological insulator (TI) by breaking time-reversal symmetry. A magnetic exchange field perpendicular to the TI plane breaks the symmetry protecting the gapless surface states and opens a local surface mass gap, switching the surface conductance without changing the bulk band inversion.
We can use this idea, coupled with the ability to inexpensively rotate the magnetization of a magnet from in to out-of-plane using voltage-gated strain on an underlying piezoelectric, to voltage-gate the conductive surface states of a topological insulator (Fig.~\ref{fig:TImag}(a)). 
This means the TI can act as a gate-controlled SOT switching element (Fig.~\ref{fig:TImag}(b)), with a high spin Hall angle for efficient current-driven write on a storage magnet on top (Fig.~\ref{fig:TImag}(c)). 

The essential consequence of the Dirac winding, in the ideal isotropic surface model, is the spin texture
\begin{equation}
    \vec{s}(\vec{k}) = \dfrac{\hbar}{2}(\hat{z}\times\hat{k}), ~~\delta\vec{s} = \sum_k\vec{s}(\vec{k})\delta f(\vec{k})
\end{equation}
which immediately implies that shifting the Fermi surface generates a net nonequilibrium spin density through the Edelstein (inverse spin Galvanic) effect. 
Because opposite crystal momenta carry opposite spins, the equilibrium spin density vanishes. An applied electric field or heavy-metal current shifts the Fermi contour $\delta f$, creating a net spin accumulation by the Edelstein effect whose direction is fixed by the topological winding of the Dirac cone. The accumulated spin leads to the SOT torque. 

The charge-to-spin conversion is enabled by the helical surface texture associated with the topological surface state; the Edelstein response itself is not a quantized topological invariant. The ability to symmetry-gate the TI locally makes it operate as a row-column selector in a crossbar array of a processor-in-memory (PIM) bitcell \cite{morshed2025prapplied}. 
\begin{figure*}[ht!] 
  \centering
\includegraphics[width=4.83in]{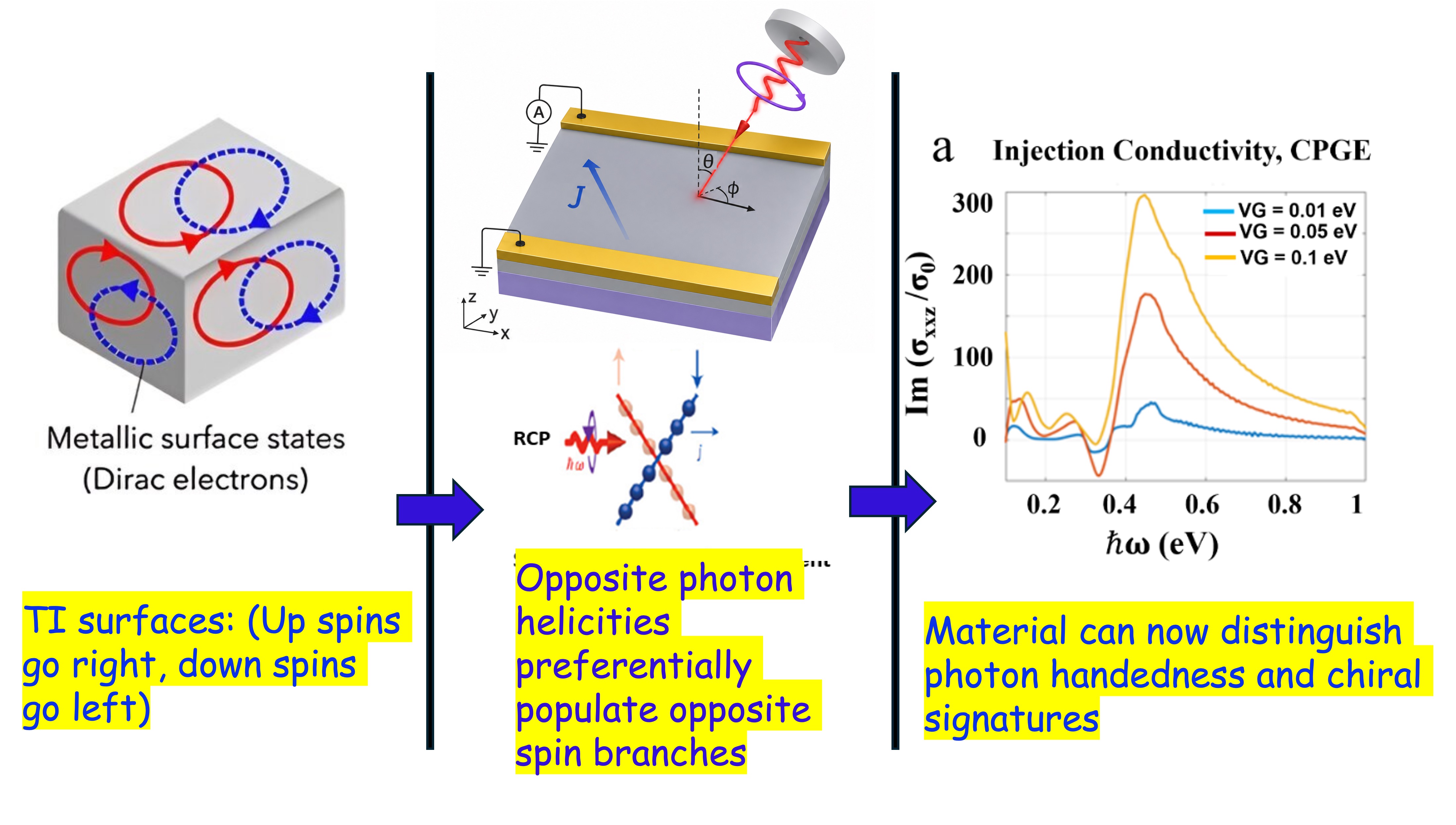} 
 \caption{Topology controlled optical selection rules. Spin-momentum locking causes opposite helicities of circularly polarized light to couple preferentially to opposite electronic spin textures. Berry curvature converts this symmetry-selective excitation into a measurable circular photogalvanic current, allowing topological materials to electrically distinguish photon handedness and optical chirality \cite{sharifpour2026}.}
 \label{fig:TI_photons}
\end{figure*}
What makes our structure unique and potentially disruptive as a PiM bitcell is the unique write mechanism associated with a FM/TI/FM layered structure – combining spin selectivity with a voltage tunable surface gap. 
Activating the TI bottom surface closes the circuit, driving the top storage magnet into one of three modes – altering its magnetism with drain bias for data write; discharging its stored magnetism for data read, its output set by its low (parallel)/high (antiparallel) resistance relative to a fixed magnet in the magnetic tunnel junction (MTJ); and executing a logic operation (e.g. bitwise AND, OR, or MAC - Multiply-And-Accumulate for synaptic sums) using a sense amplifier. Since selector and storage magnets are co-located in a vertical geometry, the structure is scalable and naturally suited for a PiM architecture that pre-processes stored data locally, all with the same vertically integrated, compact bit cell.

One challenge with the TI-based SOTRAM is that we can only deliver in-plane spins that can torque an in-plane magnet, which is hard to scale. However, this can be addressed potentially with a Weyl semimetal (WSM). The surface state based torque is based on the inverse spin Galvanic effect, but WSMs also have bulk conductive states that give a spin Hall current. Breaking certain symmetries strategically, such as a gate-driven magnetic field in the direction of current drive, shifts the Berry connection towards positive $k_y$, giving z-polarized spins traveling in the z-direction that can potentially torque an out-of-plane magnet \cite{chen2025wsm}
\begin{equation}
    j_z^{\sigma_z} = \alpha_{planar}{\cal{E}}_yB_y, ~~~ \tau_{DL} \propto j_z^{\sigma_z}
\end{equation}
reflecting the symmetry-allowed form of
the transverse spin current.
Rotating the magnetic order from y toward z suppresses the out-of-plane spin channel, acting again as a row-column selector, albeit for perpendicular MTJs.  Other spin-current components may become allowed according to the full response tensor, allowing the delivered spin symmetry to be selected by magnetic orientation.

In this example, topology and symmetry guide the conversion of electrical control of a spin-textured band structure into electrical control of the delivered spin current. In the TI, gating the surface mass selects whether an in-plane Edelstein torque is available. In a device, magnetic symmetry selects a bulk out-of-plane spin-current channel suitable for perpendicular-magnet switching.
\subsection{Topology can preferentially couple to helical photons}
Spin-momentum locking gives topological surface states helicity-dependent optical selection rules: opposite circular polarizations couple differently to oppositely oriented spin textures. A dc photocurrent nevertheless requires the appropriate crystal and device asymmetry, since inversion-related transitions otherwise cancel. The resulting imbalance appears in the second-order photoconductivity governed by the Berry geometry of the electronic bands \cite{morimoto2016,dejuan2017}
(Fig.~\ref{fig:TI_photons})
\begin{equation}
    j_a(\omega_1-\omega_2) = \sum_{bc}\sigma_{abc}(\omega_1,\omega_2){\cal{E}}_b(\omega_1){\cal{E}}_c^*(\omega_2)
\end{equation}
For photons of polarization $\alpha$ incident along the direction $(\theta,\varphi)$, the x-directed current density can be simplified as
\begin{eqnarray}
    j_x/{\cal{E}}_0^2 &=& C\sin{2\alpha} + L_1\sin{4\alpha} + L_2\cos{4\alpha} +D\nonumber\\
    C &=& -i\left[\cos{\theta}\sigma_{xy}^--\sin{\theta}\sin{\varphi}\sigma^-_{xz}+\sin{\theta}\cos{\varphi}\sigma^-_{yz}\right]\nonumber\\
    &&
\end{eqnarray}
where $C$ is the coefficient of the nonlinear circular photogalvanic effect (CPGE) \cite{connelly2024}, while the rest are linear contributions. The reduced symbols $\sigma^-_{\alpha\beta} = \sigma_{x\alpha\beta}- \sigma_{x\beta\alpha}$. The CPGE current, computed using non-equilibrium Green's functions \cite{ghosh2023,sharifpour2026}, can be interpreted schematically as
\begin{equation}
    J_{CPGE} \propto Im(\sigma_{xxz}) \propto \int_{BZ}d^3k\Omega(k)W(k)
\end{equation}
where $\Omega$ is the Berry curvature and $W$ the optical transition kernel. 
More generally, the optical matrix elements that enter the CPGE are elements of the quantum geometric tensor between resonantly connected bands. Circular polarization selects the antisymmetric component, associated with Berry curvature and helicity, while linear-polarization-resolved absorption probes symmetric combinations related to the quantum metric. Thus the photocurrent is not determined by topology alone: it is a spectrally weighted measurement of the local geometry of the spinor wavefunctions.

CPGE is symmetry-allowed in inversion-asymmetric media. In topological surface or Weyl states, Berry curvature and chiral optical matrix elements can give it distinctive magnitude, spectral structure, or—in ideal chiral Weyl systems—a quantized trace over a restricted frequency window.
Unlike the quantum Hall effect, where Berry curvature integrates to an integer Chern number, nonlinear optical phenomena weight the Berry curvature by optical transition probabilities, allowing topology to directly modulate the generated photocurrent. 

We can enhance a surface contribution with an inversion symmetry breaking vertical gate field. Breaking time-reversal symmetry through magnetic order can further enhance or reshape the CPGE by redistributing Berry curvature near the exchange gap and modifying helicity-dependent interband matrix elements. The exchange field does not merely move Weyl nodes; it redistributes the momentum-resolved Berry curvature and spin-textured optical matrix elements so that contributions previously related by crystal symmetry (e.g. tetragonal $D_{2d}$)
operations no longer cancel. 

The role of topology here is to make photon helicity electrically distinguishable through the geometry of spin-textured electronic states. Unlike quantum Hall, the resulting photocurrent is spectrally weighted by optical matrix elements, but its handedness sensitivity provides a direct route to compact chiral and polarimetric sensing.

\section{Conclusion}
Electronics engineers manipulate charge density and electric fields to build semiconductor logical gates. Topological electronics utilizes the allowed continuity of wavefunctions through symmetry and geometry to attempt the same. The challenge lies in translating these geometric constraints into scalable materials, interfaces and contacts, and manufacturing technologies.

The historical development of condensed matter has largely treated graphene, topological insulators, Weyl semimetals and magnetic skyrmions as separate subjects. We argue instead that they are manifestations of a common SU(2) geometry. Once viewed through that lens, topology ceases to be merely a mathematical invariant and instead becomes a design language for engineering. Different physical realizations constrain different quantities like spins, pseudospins and polarizations, all emerging from geometrically constrained two-component textures whose winding, symmetry and continuity govern the allowed response. This unified viewpoint suggests that future quantum materials should be classified not only by chemistry or crystal symmetry, but also by the symmetry of the underlying two-component wavefunction and the device functionality enabled by its topology.

\indent {\it{Acknowledgment}}
We acknowledge funding support by the DARPA TEE and the NSF CISE 2504227 grants, and useful discussions with Philip Kim, Cory Dean, Geoffrey Beach, Andrew Kent, Mark Stiles, Joseph Poon, Supriyo Bandyopadhyay, Patrick Taylor, Mircea Stan and George deCoster. 
\\
\bibliography{refs}
\bibliographystyle{apsrev4-2} 
\end{document}